\documentclass[aps,pre]{revtex4}

\usepackage{graphicx}
\usepackage{amssymb,amsfonts,amsmath,color}

\begin{document}

\title{Theory of time-averaged neutral dynamics with environmental stochasticity}

\author{Matan Danino and Nadav M. Shnerb}

\affiliation{Department of Physics, Bar-Ilan University,
Ramat-Gan 52900, Israel.}

\begin{abstract}
\noindent Competition is the main driver of population dynamics, which shapes the genetic composition of populations and the assembly of ecological communities. Neutral models assume that all the individuals are equivalent and that the dynamics is governed by demographic (shot) noise, with a steady state species abundance distribution (SAD) that reflects a mutation-extinction equilibrium. Recently, many empirical and theoretical studies emphasized the importance of environmental variations that affect coherently the relative fitness of entire populations. Here we consider two generic time-averaged neutral models, in both the relative fitness of each species fluctuates independently in time but its mean is zero. The first (model A) describes a system with local competition and linear  fitness-dependence of the birth-death rates, while in the second (model B) the competition is global and the fitness dependence is nonlinear. Due to this nonlinearity, model B admits a noise-induced stabilization mechanism that facilitates the invasion of new mutants. A self-consistent mean-field approach is used to reduce the multi-species problem to two-species dynamics, and the  large-$N$ asymptotics of the emerging set of Fokker-Planck equations is presented and solved. Our analytic expressions are shown to fit the SADs obtained from extensive Monte-Carlo simulations  and from numerical solutions of the corresponding master equations. \vspace{2cm}

\end{abstract}

\maketitle

\section{Introduction}

Neutral models play a central role in the theoretical analysis of population genetics and community ecology \cite{kimura1985neutral,hubbell_book,azaele2016statistical}. These models neglect the details of interspecific interactions and emphasize the role of stochastic processes as key drivers of abundance variation and species diversity. Deterministic factors (like selection, niche partitioning and species specific interactions) are not included in the model. Instead, one considers a (usually, zero-sum) competition between types (species, strains, alleles etc.) where all the individuals are functionally equivalent ('neutral'). The structure of a community, i.e., the commonness or rarity of different species, reflects the inherent stochasticity of the underlying birth-death process, while the corresponding birth-death rates are species independent and are fixed in time.

A two-species competition of this kind is described by the classical voter model \cite{liggett2013stochastic} that leads, inevitably, to the extinction of one of the species and to fixation by the other. When the model allows for mutation/speciation events the system may reach a steady state that reflects the balance between mutations and extinctions. Quantities like the species abundance distribution (SAD, aka site frequency spectrum) and the mean species richness (SR) may then be calculated as a function of the model  parameters~\cite{maritan1,azaele2016statistical}.  The ability of these SADs to account for empirically observed species abundance distributions in many high-diversity assemblages~\cite{volkov2005density,TREE2011,ter2013hyperdominance} is considered as the main success of the neutral model of biodiversity.

Despite their great influence, some aspects of the traditional neutral models are problematic. In particular, these models assume that the  dynamics is driven by a \emph{stationary} birth-death process. Under this assumption,  variations in abundance of a species reflect the cumulative effect of the \emph{uncorrelated} reproductive success of all its individuals.  In such a binomial process both the per-generation population variance and the time to extinction (in generations) scale linearly with the population size. In contrast, many empirical analyses  show  that the magnitude of temporal abundance variations is much higher \cite{dornelas2006coral,leigh2007neutral,kalyuzhny2014niche,
chisholm2014temporal,hekstra2012contingency,kessler2015neutral},
that the scaling of population variance with population size is superlinear \cite{kalyuzhny2014temporal,
chisholm2014temporal} and that the  rate of changes in species composition is much faster than the predictions of the neutral model \cite{feeley2011directional,kalyuzhny2015neutral}.

The simplest solution to that problem is \emph{environmental stochasticity}  \cite{lande2003stochastic} (also known as fluctuating selection~\cite{bell2010fluctuating,bergland2014genomic}, temporal niches etc.): a time-varying environment may  alter the demographic parameters (such as growth and mortality rates) and the competitive ability of an entire population, so the reproductive success (say, the average number of offspring) of all the conspecific individuals increases or decreases in a correlated manner. Accordingly, population variance scales with $n^2$, where $n$ is the population size. The stochastic process is no longer stationary, and at any given time some species are superior and others are inferior. The model may still be considered as "neutral"~\cite{loreau2008species} if the time averaged fitnesses of all species are equal (time-averaged neutrality, \cite{kalyuzhny2015neutral}).  Numerical and empirical analyses suggest that time-averaged neutral models of this type may explain both static and dynamic patterns in ecological communities \cite{kalyuzhny2015neutral,fung2016reproducing}. These observations raise the need for an analytic solution for time averaged neutral models.

 A few versions of the two-species time-averaged neutral model  were considered recently (sometimes in the context of the speed of evolution \cite{cvijovic2015fate,danino2017environmental}), and quantities like the chance of fixation and the time to fixation were calculated \cite{assaf2013cooperation,danino2016effect,danino2016stability,hidalgo2017species,danino2017fixation,wienand2017evolution}. Other works dealt with the dynamics of a single species under environmental variability, trying to infer the SAD of the corresponding multi-species neutral model from the results \cite{huerta2008population,kessler2014neutral,fung2017species}.  Here we present a solution for the species abundance distribution in a multi-species time-averaged neutral model, where the process of species extinction is compensated by the introduction of new types via mutation/speciation events.   Our results are given in terms of the chance for mutation, the strength of demographic noise and the amplitude of environmental variations, the relevant definitions are summarized in  Table \ref{table1}.

 \begin{table}[h]
\begin{center}
\caption {Glossary} \label{table1}
    \begin{tabular}{ | l |  p{10cm} |}

    \hline
    Term  &  Description \\ \hline
    $N$ &  number of individuals in the community. The strength of demographic noise scales like $1/N$.  \\ \hline
    $\nu$ &  The chance of mutation/speciation (per birth). \\ \hline
    $\theta \equiv N \nu$ &  The fundamental biodiversity  number. Mean number of mutations per generation. \\  \hline
    $\delta$ & correlation time of the environment, measured in  generations.\\
    \hline
     $\gamma$ & the amplitude of the fitness fluctuations.  \\ \hline
     $g$ & $\gamma^2 \delta/2$, the strength of environmental stochasticity. \\ \hline
     $G \equiv N g$ & The ratio between environmental stochasticity and demographic noise. \\ \hline
     $\nu/g = \theta/G$ & the ratio between mutation load and environmental stochasticity.  \\ \hline
    \end{tabular}

\end{center}

\end{table}

Technically speaking, neutral models are easier to solve since the multi-species problem may be reduced to a set of (identical) single species problems \cite{etienne2007zero}. The abundance of a focal species, $n$, and the size of the community $N$, fully determine the transition rates of this focal species, since demographic equivalence implies that the partitioning of the $N-n$ individuals among all other species is irrelevant. This feature is lost when environmental variations are taken into account, as the instantaneous fitness of all other individuals does affect the focal species.  We will show that, in high diversity assemblages, this obstacle may be overcome using an effective medium theory that becomes even simpler in the large $N$ limit.

To pave the way for this analysis, we  will consider first a two-type, one-way mutation model with environmental stochasticity. In this model the state (abundance and fitness) of the focal species determines unambiguously the state of the whole system, so the analysis is relatively easy. Then we will show that the full, multi-species model may be reduced (with appropriate modifications) to the two species case and, using this feature, we obtain the required SADs.

To facilitate the discussion, we introduce three appendices in which  technicalities are introduced and discussed. Appendix \ref{apa} explains, using a simple example, the transition from the master equation to the Fokker-Planck equation with a particular emphasis on the boundary conditions. The corresponding calculations for the two-species, one-sided mutation case are presented in Appendix \ref{apb}, and the relevant modifications that allow us to solve the time-averaged neutral model are discussed in Appendix \ref{apc}.

\section{Model A and model B: Environmental stochasticity and noise induced stabilization}

In this section we would like to provide a few basic insights regarding the effect of environmental variations, and in particular to make a distinction between microscopic models that lead to noise-induced stability and those that do not support this feature. Our two examples here involve global and local competition; we first present these models with pure demographic noise, where they lead to the same outcome, then we will explain their different behaviour in fluctuating environment.

As an example of local interactions (model A), one may imagine two populations that leave together on, say, an island. Individuals are wandering around, looking for food, mate or territory. An encounter between two individuals may lead to a struggle in which only one of them wins the desired goods and increases its chance to survive and to reproduce. In a zero-sum game of this kind two individuals are chosen at random from the entire community for a duel, the loser dies and the winner produces a single offspring. If one consideres a two species community of size $N$, where the fraction of one species is $x = n/N$, the chance for an interspecific duel is $2x(1-x)$. In a neutral model without environmental variations all individuals have equal fitness all the time, so the chance to win a duel is always $1/2$. Accordingly, the chance of a population to grow or to decrease by one individual after a single elementary event (a duel) is equal, $x(1-x)$.

To present a model with global competition (model B), let us consider a forest. Adult trees spread seeds all around and we assume that the  dispersal length is much larger than the size of the forest, so the composition of the seed bank at each location reflects the abundance of the corresponding species in  the forest. When an adult tree dies it leaves a gap and one local seed is chosen to capture it. If the model is neutral the chance of each species to recruit the gap is proportional to its relative abundance. Hence, the abundance of $x$ will grow by one tree with probability $x(1-x)$ (an adult tree from another species has been chosen to die w.p. $(1-x)$ and this species won the gap w.p. $x$) and will shrink by one tree with the same probability.

Accordingly, when the environment  is fixed and the dynamics is purely neutral, the local competition  model (A) and the global competition model (B) are translated to the same stochastic process (the voter model) and lead to the same dynamics. However, this feature is lost when  environmental fluctuations do affect the relative fitness of different species, even if the averaged fitness differences  vanish.

To model environmental stochasticity we begin with a two species game, and later on we will extend the definition to the general case. Focusing on a specific species with relative abundance $x$, in model A the chance of an interspecific duel is  $2x(1-x)$. We will define the  fitness of this species (with respect to its enemy) via the chance to win such a duel,
\begin{equation}
P_{win} = \frac{1}{2} + \frac{\gamma(t)}{4},
\end{equation}
where $\gamma(t)$ measures its relative (log) fitness. The focal species  mean population satisfies (time is measured in generations, $N$ elementary events in each generation),
\begin{equation}
\dot{x} = \gamma x (1-x),
\end{equation}
or $\dot{z} = \gamma z$, where $z \equiv x/(1-x)$. Accordingly, if $\gamma$ is fixed in time the focal species abundance grows (when $\gamma$ is positive) or decays (for $\gamma<0$) and the focal species reaches fixation ($z>1-1/N$) or extinction ($z<1/N$) on ${\cal O}(\ln N)$ timescales. Our interest here is in a time-averaged neutral model where $\gamma(t)$ has zero mean. In that case $z(t)$ performs a simple unbiased random walk without any stabilizing force.

In model B the role of environmental variations is not so simple. If the fitness affects the chance of recruitment but death occurs randomly, the chance of the focal species to increase its abundance is equal to the chance that a tree from another species dies, $1-x$, times the chance of the focal species to win the empty slot,  $x e^\gamma /(1-x+xe^\gamma)$. The focal species shrinks if one of its individuals was chosen to die (with probability $x$) and the other species wins the competition with probability  $(1-x)/(1-x+xe^\gamma)$. Accordingly, $x$ satisfies,
\begin{equation}
\frac{dx}{dt} = \frac{x (1-x) e^\gamma}{1-x + x e^\gamma}-\frac{x (1-x)}{1-x + x e^\gamma} \approx \gamma x(1-x) + \frac{\gamma^2}{2} x(1-x)(1/2-x),
\end{equation}
where the last term comes from a second order expansion in $\gamma$. Unlike model A, here the nonlinear dependence of the chance to win on $\gamma$ leads to a second, ${\cal O}(\gamma^2)$ term, that by itself tends to stabilize the coexistence point at $x=1/2$. Of course this term is much smaller than the first, ${\cal O}(\gamma)$ term, so under fixed environmental conditions the focal species still shrinks or grows exponentially. However, when $\gamma(t)$ fluctuates around zero the ${\cal O}(\gamma)$ term averages out while the ${\cal O}(\gamma^2)$ terms add up, so (at least when the rate of variations is fast enough) the stochasticity tends to stabilize the coexistence point.

 The difference between model A and model B is most evident when the environmental fluctuation are extremely rapid, e.g., when $\gamma$ is picked at random after each elementary (birth-death) event. Model A reduces, in this case, to its purely demographic limit: instead of choosing the winner by a single toss of a coin one first picks the weather and then the winner, but the end result is a chance of $1/2$ to win any elementary competition. In contrast, in model B the stabilizing effect of the environment reaches its maximum  strength in this rapid fluctuations limit where the ${\cal O}(\gamma)$ terms cancel each other more efficiently.

 The stabilizing effect of environmental variations in models with nonlinear fitness dependence (like our model B) was pointed out by Chesson and coworkers while ago \cite{chesson1981environmental,chesson1994multispecies}. Technically, model B considered here is very close to Chesson-Warner lottery game. However, as discussed in \cite{danino2016effect}, the lottery game has no demographic noise, so it does not allow for extinction events and one cannot analyze the properties of a community in which the biodiversity reflects an extinction-speciation equilibrium.

\section{An individual-based  two-type model with environmental stochasticity and  one-way mutation} \label{sec2}

As explained above we shall start our analysis, in this section, with a two-species game, and then (in section \ref{sec4}) extend the treatment to the full problem. In these two sections we begin with model A, and then consider model B.

\subsection{Model A} \label{sec2a}

Let us consider a system of $N$ individuals with two species (types), $A$ and $B$. As in \cite{karlin1981second} (p. 208) no mutation of $B$ to $A$ is allowed, while an  offspring of type A may mutate to become a B-type.

 In each elementary event two  individuals are picked at random,  the winner reproduces and the loser dies.  If a $B$-type wins, the offspring is also a $B$. If an $A$ wins, the offspring is an $A$ with probability $1-\nu$ and mutates to be a $B$-type  with probability $\nu$.

Accordingly, in a system of $N$ individuals with $n$ A-types and $N-n$ B-types, the only absorbing state is $n = 0$. In this section we assume that, very rarely, a new A-type individual arrives (say, as an immigrant) and then the game is played again until the $A$ species goes extinct (this happens before the next immigration event). Our aim is to calculate $P_n$, the probability to find the system with $n$  A-types, \emph{conditioned} on the existence of A in the system (i.e., not including the periods between extinction  and recolonization events).

 As explained, this  process takes place via a series of duels. In case of an interspecific duel $A$ wins with probability $P_{win}$ (to be defined below) and  $B$ wins with probability $1-P_{win}$. The possible outcomes of all kinds of duels are summarized by (here the expressions above the arrows are probabilities, not rates),
 \begin{eqnarray} \label{eq1}
 B+B \xrightarrow{1} 2B \qquad A&+&A  \xrightarrow{1-\nu} 2A  \qquad A+A  \xrightarrow{\nu} A+B \nonumber \\
 A+B \xrightarrow{1-P_{win}} 2B \qquad A&+&B \xrightarrow{P_{win}(1-\nu)} 2A \qquad A+B \xrightarrow{\nu P_{win}} A+B.
 \end{eqnarray}

 To fully characterized the process, $P_{win}$ should be specified.  We define $P_{win}$ via
 \begin{equation}\label{eq2}
      P_{win} = \frac{1}{2} + \frac{s_A - s_B}{4},
\end{equation}
 where $s_A$ ($s_B$) is the logarithmic fitness of the $A$ ($B$) type. Without loss of generality we can set $s_B=0$. Under environmental variations $s_A$ (hence  $P_{win}$) is time-dependent, but to keep time-average neutrality its mean has to be zero. Clearly, the main characteristics of such environmental fluctuations are their amplitude and their correlation time. Here we assume a dichotomous (telegraphic) noise  such that  $s_A= \pm \gamma$, so half of the time $P_{win} = 1/2+\gamma/4$ (the plus state of the environment) and half of the time $P_{win} = 1/2 - \gamma/4$ (the minus state). Both white Gaussian noise and white Poisson noise can be recovered from the dichotomous noise by taking suitable limits~\cite{ridolfi2011noise}, so the results obtained here are quite generic.

 Time is measured in units of generations, where a generation is defined as $N$ elementary duels. After each elementary duel the environment switches (from $\pm \gamma$ to $\mp \gamma$) with probability $1/(N \cdot\delta)$, so the sojourn times of the environment (measured in generations) are geometrically distributed with mean  $\delta$.

 At this point the model is fully specified. A full list of the transition probabilities is given in Appendix \ref{apb}, Eq. (\ref{eqb2}). Using that, one may write down the corresponding set of master equations (\ref{eqb1}). In Appendix \ref{apb} we show how to derive, from this exact master equation, an effective Fokker-Planck equation for $P(x)$, the chance (averaged over time, including plus and minus periods) to find the system with  $n \equiv Nx$ $A$-type individuals, satisfies
 \begin{equation} \label{eq3}
\left[ x(1-x) \left(1 + Gx(1-x) \right) P(x) \right]'' - \left[ ( Gx(1-x) (1-2x) - \theta x) P(x) \right]' = 0.
\end{equation}
Here tags are derivatives with respect to $x$  and $G \equiv  N \delta  \gamma^2/2$ is the  ratio between the effective strength of the environmental stochasticity, $g = \gamma^2 \delta /2$, and $1/N$, the strength of the demographic noise. The fundamental biodiversity  number  $\theta = N \nu$ is a measure of the population mutation rate (mutation load per generation).

Solving for $P(x)$ with the appropriate boundary conditions (see Appendix \ref{apa} and \ref{apb}, where we explain this subtle issue), one obtains,

\begin{equation} \label{eq4}
P(x) = C \frac{(1-x)^{\theta}}{x(1-x) [1 + G x (1-x) ]^{\theta /2}} \left[ \frac{1-(1-2x)\sqrt{\frac{G}{4+G}}}{1+(1-2x)\sqrt{\frac{G}{4+G}}} \right]^{\frac{\theta}{2}\sqrt{\frac{G}{4+G}}}
\end{equation}
where $C$ is a normalization factor.
To provide a background for later discussions, let us consider a few features of the solution (\ref{eq4}).

\begin{figure}
\includegraphics[width=8cm]{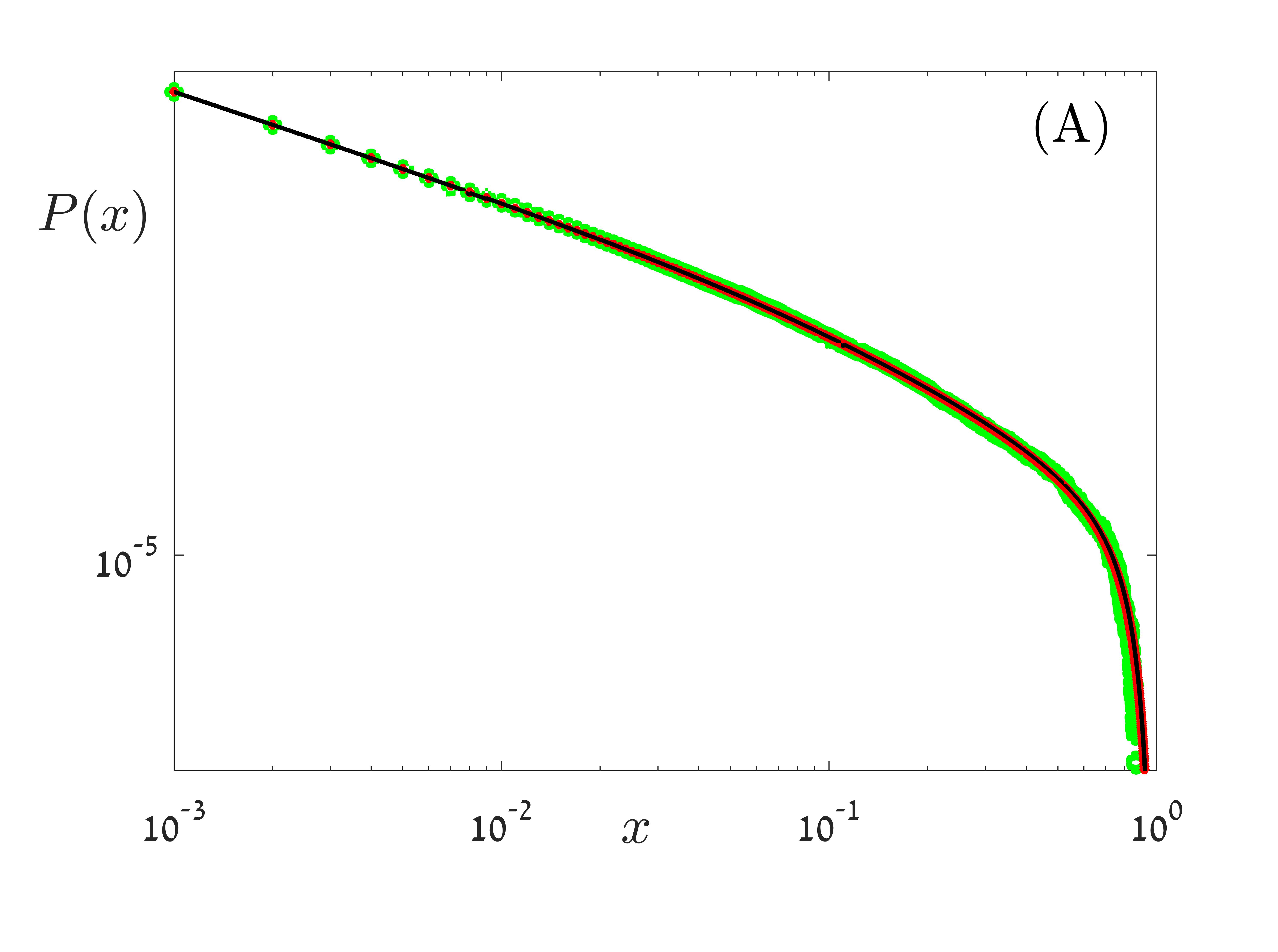}
\includegraphics[width=8cm]{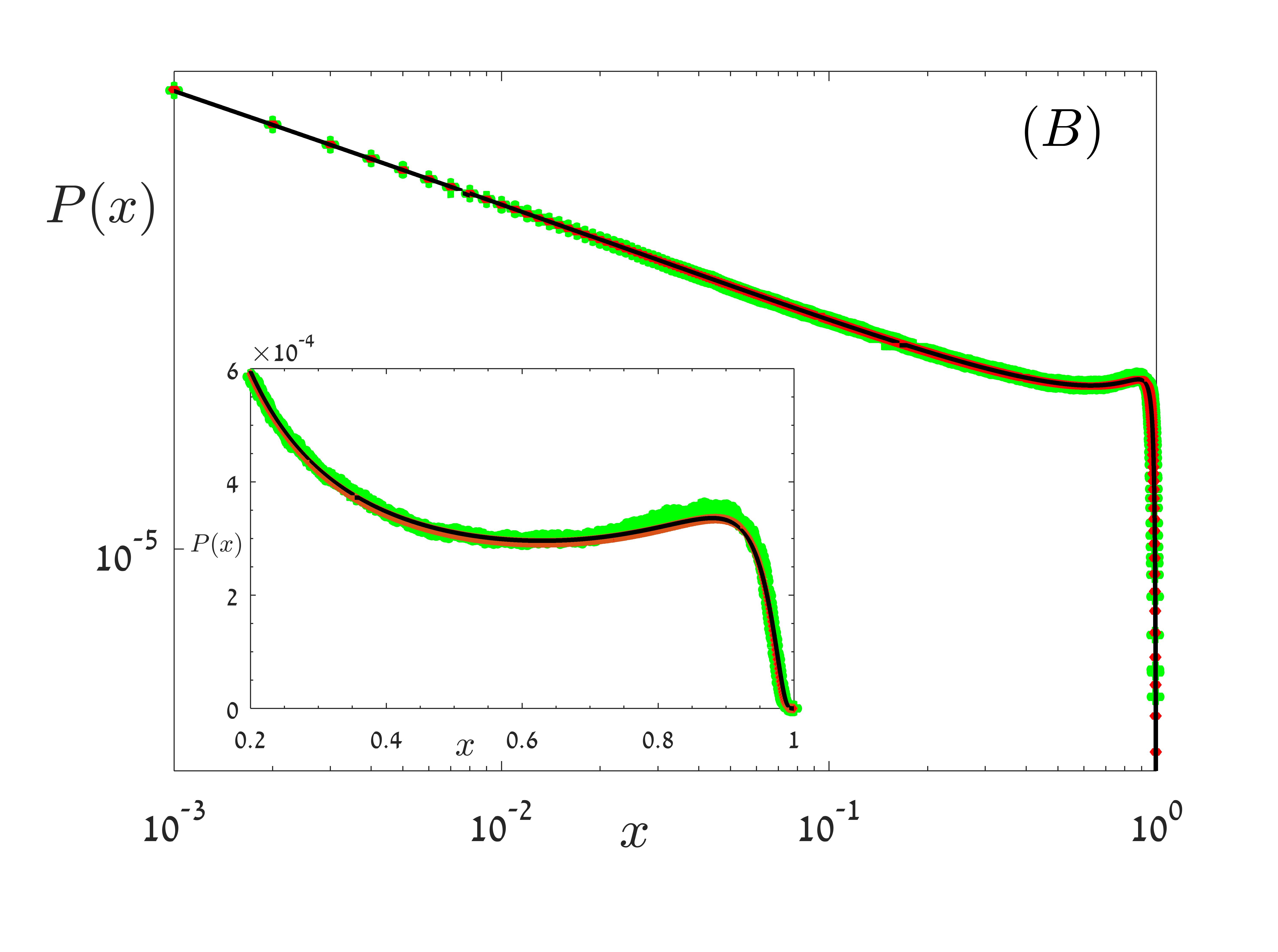}
\caption{$P(x)$, the chance of finding the $A$-type at relative abundance $x$, is plotted for a  two competing species system with  one-sided mutation, environmental stochasticity and demographic noise. In both figures $N = 1000$, $\nu = 0.01$ and the main parts are plotted using a double logarithmic scale. Results shown include those obtained from a Monte-Carlo simulation (green circles), numeric solutions for the steady state of the master equations (\ref{eqb1}-\ref{eqb3}) (red diamonds) and the analytic prediction of Eq. (\ref{eq4}) (black line). In panel (A) the results are depicted  for $\delta = 0.5$ and $\gamma = 0.2$, such that $\nu =g$. In panel (B) $\delta = 1.25$ and $\gamma = 0.4$ so $\nu = 0.1g$. As discussed in the main text, when $\nu < g/2$ there is a peak at high values of $x$. To emphasize this peak we have added an inset where the same results are shown using a linear scale. The fit between the three curves is quite good. } \label{fig1}
\end{figure}

\begin{enumerate}

\item When $G \to 0$ (no environmental stochasticity) we have a model with mutations and demographic noise. In that case, $P(x)$ obtained from our two species model is simply,
\begin{equation} \label{eq5}
P(x) = C \frac{(1-x)^\theta}{x(1-x)},
\end{equation}
i.e., the Fisher log-series that converges to  $e^{-\theta x}/x$ when $\theta \gg 1$. In this case the two-species model yield the SAD of the neutral model, since there is no real difference between the two. Every species in the neutral model emerges via mutation/speciation and goes extinct because of demographic noise, so the average over colonization-extinction periods that yields $P(x)$ is the same as the average over different  species that yields the SAD of the neutral model. As we shall see below, when environmental stochasticity kicks in, $P(x)$ of the two species model differs from the SAD of the neutral model.

 When $\theta \ll 1$ the expression for $P(x)$ in Eq. (\ref{eq5}) reduces to $[x(1-x)]^{-1}$, since in that case the system spends most of its time close to the fixation/extinction points in a symmetric manner.

\item For strong environmental stochasticity, i.e.,  when $G \gg 1$, one may use the approximation $\sqrt{G/(G+4)} \approx 1-2/G$. When this expression is plugged into Eq. (\ref{eq4}) and constants are absorbed into the normalization factor one obtains,
\begin{equation} \label{eq6}
P(x) = C (1-x)^{ \frac{\nu}{g} -1} \left( \frac{(1+Gx)(1-x)}{1 + G x (1-x)} \right)^{\theta /2} \frac{(1+Gx)^{-\nu/g}} {x}.
\end{equation}

 When all parameters are kept fixed  and $x$ decreases such that  $Gx \ll 1$ and $\theta x \ll 1$ (which implies, of course, also $x \ll 1$), the dynamics is purely demographic and Eq. (\ref{eq6}) reduces to,
  \begin{equation} \label{eq61}
  P(x) \sim \frac{1}{x}.
  \end{equation}

 On the contrary, in the region where the demographic noise in negligible,  $Gx \gg 1$,
\begin{equation} \label{eq7}
P(x) \sim  \frac{(1-x)^{\frac{\nu}{g}-1}}{x^{\frac{\nu}{g}+1}} \exp \left(-\frac{\theta x/2}{1+Gx(1-x)} \right).
\end{equation}
\begin{itemize}
\item When $\nu > 2g$, the exponent in (\ref{eq7}) truncates $P(x)$ at
\begin{equation}
x_c = \frac{1}{N(\nu/2-g)}.
\end{equation}
In the large $N$ limit  $x_c \ll 1$ so  $(1-x)^{\nu/g-1} \approx 1 $. Accordingly, $P(x)$ looks like $1/x$ for $x \ll 1/G$, like $x^{-1-\nu/g}$ in the narrow region $1/G \ll x \ll 1/(\theta/2-G)$, and decays to negligible values above this point. The intermediate power-law regime disappears when $\nu>4g$, where $P(x)$ takes the general form of the Fisher log-series with an effective mutation rate which is half of the bare mutation rate, plus some modifications due to $G$ in the tail of the distribution.

\item When $\nu=2g$ the exponential term in Eq. (\ref{eq7}) still provides a cutoff, now at $x_c \sim 1/\sqrt{G}$. Since $x_c$ is still microscopic in the large $N$ limit, the intermediate  power-law $x^{-3}$ is valid in the region $1/G \ll x \ll 1/\sqrt{G}$.

\item If $\nu<2g$, the exponential cutoff point becomes $N$-independent,
\begin{equation}
x_c = 1-\frac{\nu}{2g},
\end{equation}
so,
\begin{equation} \label{eq7a}
P(x) \sim  \frac{(1-x)^{\frac{\nu}{g}-1}}{x^{\frac{\nu}{g}+1}} e^{-\frac{\nu}{2g(1-x)}}.
\end{equation}
Below $x_c$, the behavior is determined by the pre-exponential factor. If $g<\nu<2g$, this factor decays monotonously with $x$, so one observes two power-laws with an Arrhenius  truncation above $x_c$. In the region $\nu<g$ the prefactor grows with $x$ above $x^*=1/2+\nu/(2g)$. When $x^*<x_c$, i.e., $\nu < g/2$, $P(x)$ admits an observable peak at finite $x$, as demonstrated in Figure \ref{fig1}B.
    \end{itemize}

\end{enumerate}

The adequacy of Eq. (\ref{eq4}) and the different behaviors of $P(x)$   are demonstrated in Figure \ref{fig1}. The analytic predictions are shown to fit the outcomes of Monte-Carlo simulations  and  the numerical solutions of the master equation.  As expected, when $g > \nu$ a peak appears close to $x=1$.

\subsection{Model B} \label{sec2b}

In model B, each elementary step begins with the death of a randomly chosen  individual, so death-probability is  fitness independent. In our one-sided mutation game, with probability $\nu$ the gap is recruited by a $B$ type individual. With probability $1-\nu$ the chance of each species to capture the vacancy is proportional to its abundance, weighted by its fitness. Accordingly, if the relative logarithmic fitness of A-type is $\gamma$ and its fraction is $x$, its chance to increase its population by one comes from events where a B individual was chosen to die (w.p. $1-x$) and no mutation happens, so the transition probabilities are
\begin{equation}
W_{n \to n+1} = (1-\nu)\frac{(1-x)x e^\gamma}{1-x+xe^\gamma}, \qquad W_{n \to n-1} =  x \left(\nu + (1-\nu)\frac{(1-x)}{1-x+xe^\gamma}\right).
\end{equation}
Under dichotomous environmental stochasticity after each step the system switches from $\pm \gamma$ to $\mp \gamma$ with probability $1/(N\delta)$.

Implementing the same method used for model A, one finds the corresponding Fokker-Plank equation,
\begin{equation} \label{eq3modelB}
\left[ x(1-x) \left(1 + Gx(1-x) \right) P(x) \right]'' - \left[ ( G \eta x(1-x) (1-2x) - \theta x) P(x) \right]' = 0.
\end{equation}
where $\eta \equiv 1+1/\delta$. The only difference between this Fokker-Planck equation and the  equation for model A (Eq. \ref{eq3}) is the innocent looking factor $\eta$. However, this may lead to a substantial modification of the results. In model A, the deterministic bias towards $x=1/2$, represented by the $G x(1-x)(1-2x)$ in the convection term, is balanced by the decrease in the diffusion rate close to the edges, related to the factor $Gx^2(1-x)^2$ in the diffusion term, and the two phenomena cancel each other exactly in the steady state \cite{huerta2008population}. Since $\eta >1$, the attraction towards $1/2$ is dominant in model B, hence the steady-state probability may have a peak at a finite value of $x$.

The steady state of model B turns out to be,
\begin{equation} \label{eq4modelB}
P(x) = C \frac{(1-x)^{\theta}}{x(1-x) [1 + G x (1-x) ]^{-1/\delta +\theta /2}} \left[ \frac{1-(1-2x)\sqrt{\frac{G}{4+G}}}{1+(1-2x)\sqrt{\frac{G}{4+G}}} \right]^{\frac{\theta}{2}\sqrt{\frac{G}{4+G}}} =  [1 + G x (1-x)]^{1/\delta} \cdot P_{\rm{Model \  A}}
\end{equation}
Clearly, the extra term has a maximum at $x=1/2$ and the peak becomes more pronounced when $\delta$ decreases, as expected.

\begin{figure}
\includegraphics[width=8cm]{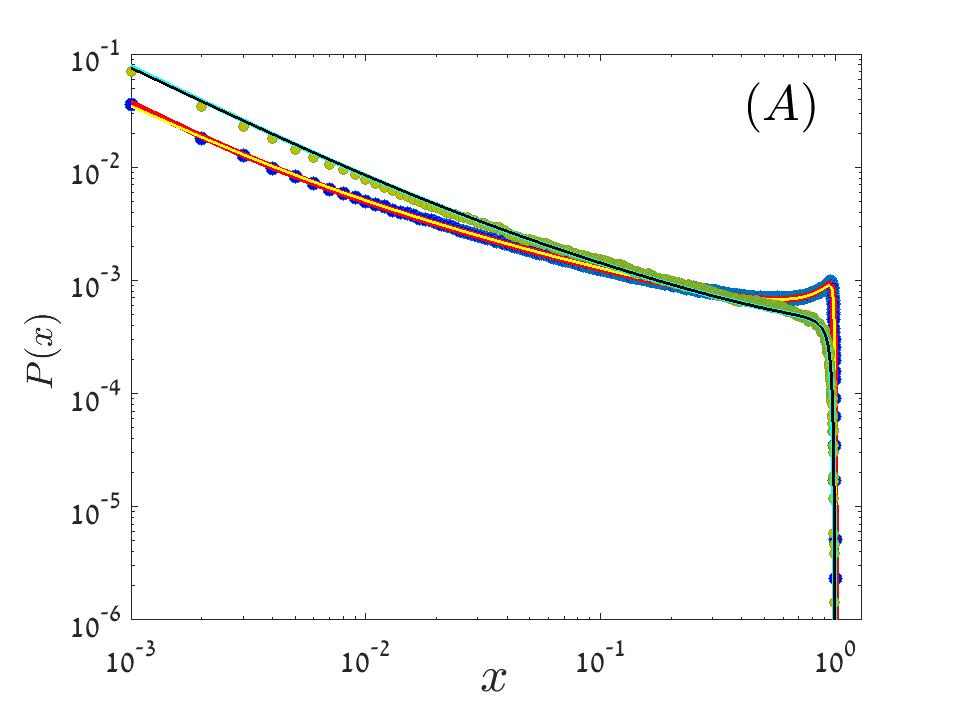}
\includegraphics[width=8cm]{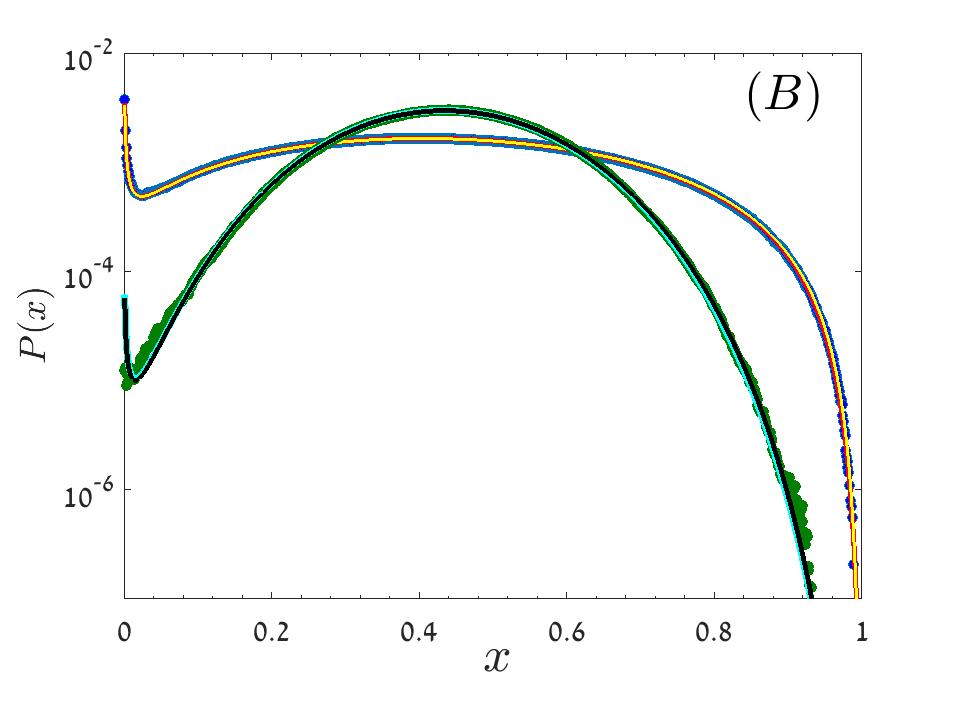}
\caption{$P_{\rm {model \  B}}(x)$, the chance of finding the $A$-type at relative abundance $x$, is plotted for a system with two competing species with one-sided mutation, environmental stochasticity and demographic noise. In both figures $N = 1000$, $\nu = 0.005$.  Results shown include those obtained from a Monte-Carlo simulation (filled circles), numeric solution for the steady state of the master equations and the analytic prediction of Eq. (\ref{eq4modelB}) (lines with different colors). In panel (A) (plotted using a double logarithmic scale) the results are depicted  for  large value of $\delta$, $\delta = 2$, so the outcomes imitate those obtained for model A, in particular the two power-laws when $\gamma =0.2$ (green circles) and the peak close to $x=1$ when $\gamma = 0.4$ (blue circles).  In panel (B) (where the scale we have used is semi-logarithmic) $\gamma = 0.4$ while $\delta =0.4$ for the blue circles and $\delta = 0.1$ for the greens. Since $\delta$ is small, the peak at $x=1/2$ is pronounced, and it becomes even steeper as $\delta$ decreases. In all these graphs the fits are good, and one can hardly distinguish between the numeric solution for the steady state and the analytic expression (\ref{eq4modelB}).  } \label{fig1modelB}
\end{figure}

Therefore,  model B is richer than model A: for  $\delta \gg 1$ $P(x)$ in model B may yield the same behaviors described above, such as a truncated power-law or a peak close to $x=1$.  However, when $\delta \ll 1$  the stabilizing force is strong and the probability develops a peak close to $x=1/2$. These  behaviors are demonstrated in Figure \ref{fig1modelB}

\section{A multi-species, time-averaged neutral model} \label{sec4}

 Having solved the problem of a two-species system with environmental stochasticity and one-sided mutation, we return to the main goal of this paper: the attempt to find  the SAD  of a neutral model with both demographic and environmental stochasticity. In this model the system may support many species, and each of these species is characterized by its abundance $n$ and by its instantaneous fitness.

Without environmental noise the dynamics of every focal species in a neutral system is identical to the dynamics of type $A$ in the two-species one-sided mutation model considered in the previous section.  Accordingly, as demonstrated in the last section,  in that case the function $P(x)$ of the two-species model is proportional, up to a normalization constant,  to the SAD of the multi-species neutral model (defined also as $P(x)$, but now it is the probability that a randomly picked species has abundance $x$). When the environmental variations change the relative fitness of different species this is not the case anymore. In this section we develop an effective field theory that allows us to map the neutral model to a (slightly modified) two-species system. Once this goal is achieved, we can solve for the SAD using the techniques presented above. Again, we begin with a discussion of model A, then we consider model B.

\subsection{Model A}

As before, in each elementary step two individuals are picked randomly for a duel, and the winner is determined  with probability that depends on their relative fitness. The offspring takes the species identity of its parent with probability $1-\nu$ and becomes the originator of a new species with probability $\nu$. Unlike the two-species model, here there are no  recurrent mutations  - an offspring cannot mutate into an existing type (an infinite allele model). As a result,  the structure of the community reflects the balance between mutations and extinction events.

The environmental noise is again dichotomous: there are  two fitness state, $\pm \gamma$, and the fitness of every species jumps randomly between these two states, such that the sojourn times are distributed geometrically with mean of $\delta$ generations. The states of different species are not correlated, and the fitness of an originator of a species is chosen at random upon  its birth.  Accordingly, in this time averaged neutral model there are two types of duels: the two randomly picked individuals may have the same fitness (either plus or minus), in which case the chance of each of them to win the duel is $1/2$, or they may have different fitnesses, in which case the corresponding chances will be $1/2 \pm \gamma/2$. Unlike the two-species model considered in the last section, here two fighting individuals may have the same fitness, so the $\gamma/4$ factor above has to be replaced by $\gamma/2$ to keep the relationship between the environmental fluctuations and the demographic noise fixed.  A full specification of the model, including all the transition probabilities, is provided in Appendix \ref{apc}.

Let us consider now the dynamics of a single (focal) species. As opposed to the two-species system considered above, here when an individual of the focal species is chosen for an interspecific duel, the fitness  of its rival is  not specified uniquely by the  focal species fitness. For example, if the focal species is in the plus  state, it may compete with either an inferior or an equal individual. Therefore, to analyze the dynamics of the focal species we need an extra parameter $f_+$, the chance that its rival in an interspecific duel will be in the plus state. \emph{If} $f_+$ is a constant (time, state and abundance independent - see discussion below), then the chance of a focal species individual to win a duel, when the focal species is in the plus state, is,
\begin{equation}  \label{eq8}
q  = f_+ \frac{1}{2} + (1-f_+) \left(\frac{1}{2} + \frac{\gamma}{2} \right) = \frac{1}{2} + \frac{\gamma}{2} (1-f_+).
\end{equation}
and when $f_+ = 1/2$ the dynamics reduces to the two-species model considered above.

The introduction of the constant $f_+$ allows us to implement the method presented in the last section to the dynamics of a focal species in the time-averaged neutral model.   In Appendix \ref{apc} we show that, in this case, $P(x)$ of an arbitrary focal species (and hence the SAD of the model) satisfies,
\begin{eqnarray} \label{eq9}
\left[ x(1-x)\left(1 + G x(1-x)\right) P(x) \right]'' - \left\lbrace \left[ Gx(1-x)(1-2 x) + N \gamma  x(1-x) (1-2 f_+) - \theta x\right]P\right\rbrace ' = 0.
\end{eqnarray}
The solution of this equation is:
\begin{equation} \label{eq10}
P(x) = C \frac{(1-x)^{\theta}}{x(1-x) [1 + G x (1-x) ]^{\theta /2}} \left[ \frac{1-(1-2x)\sqrt{\frac{G}{4+G}}}{1+(1-2x)\sqrt{\frac{G}{4+G}}} \right]^{\left(\frac{\theta}{2}-\zeta\right)\sqrt{\frac{G}{4+G}}},
\end{equation}
where $$\zeta \equiv -\frac{2}{\gamma \delta}(1-2 f_+).$$

In general $f_+$  may depend on the abundance of the focal species. However, when the abundance of each species is only a tiny fraction of $N$ (which is the case when the system supports many species, see  below) one may expect it to be independent of the details of the state of the system. Our numerics shows that taking $f_+$ as a constant becomes a very good approximation when $N$ is large. In fact, $f_+$ turns out to be independent of the abundance and the state (plus/minus) of the focal species, but it fluctuates in time. Since the transition rates are linear in $f_+$, their average depends only on its mean, $\overline{f}_+$.

 Given that, we can obtain a closed form for the species abundance distribution by calculating $\overline{f}_+$ as a function of the system parameters. If all species are "microscopic" ($n \ll N$)   $\overline{f}_+$ has to be, more or less, the fraction of individuals in the plus state, so it satisfies the self-consistency equation,
\begin{equation} \label{eq11}
\overline{f}_+ = \frac{1}{\overline {x}} \int_0^1 x P^+(x) \ dx
\end{equation}
where $P^+(x)$ ($P^-(x)$) is the probability that the focal species' fraction is $x$ at the plus  (minus) state and $\overline {x} = \int_0^1 x [P^+(x)+P^-(x)]$.  This, plus the relationship we  have derived from the master equations in Appendix \ref{apc} (Eq. \ref{eqc3}),
\begin{equation} \label{eq12}
\frac{\gamma \delta}{2}\left[ x (1-x) P\right]' = P^-(x)-P^+(x),
\end{equation}
leads, via integration by parts, to
\begin{equation} \label{eq13}
\zeta = -\frac{2}{\gamma \delta} (1-2\overline{f}_+) = \frac{1}{\overline{x}} \int_0^1 x(1-x)P(x) =1- \frac{\overline{x^2}}{\overline{x}}.
\end{equation}

Equations (\ref{eq10}) and (\ref{eq13}) provide a closed form for the species abundance distribution of the neutral model: the normalization constant $C$ cancels out in (\ref{eq13}), so one may use (\ref{eq13}) to determine $\overline{f}_+$ which, in turn, specifies uniquely $P(x)$.
Moreover, if $P(x)$ decays faster than $x^{-2}$, the quantity  $ \overline{x^2}/\overline{x}$ tends to zero as $N \to \infty$, so asymptotically
\begin{equation} \label{eq14}
\zeta = -\frac{2}{\gamma \delta} (1-2\overline{f}_+) \to 1.
 \end{equation}

 The same result emerges from a simple argument about the dynamics of $f_+$: when all the species are microscopic, $\dot{f}_+  = 2 \gamma f_+ (1-f_+) - f_+/\delta + (1-f_+)/\delta$, so (when $\gamma \delta \ll 1$) the steady state is $\overline{f}_+ \approx 1/2 + \gamma \delta /4$, in agreement with (\ref{eq14}).

 Given that, one may easily recognize the qualitative features of our main result, Eq. (\ref{eq10}). As in the two-species case, when all other parameters are kept fixed and $G \to 0$, the Fisher log-series distribution is recovered. When $G$ is large (\ref{eq10}) reduces to, \begin{equation} \label{eq15}
P(x) = \frac{C}{x(1-x)}  \left[ \frac{(1-x) \left(\frac{1}{G}+x \right)}{1+Gx(1-x)} \right]^{\theta/2} \left[ \frac{ \left(\frac{1}{G}+x \right)}{1-x} \right]^{-\zeta - \nu/g}.
\end{equation}

\begin{figure}
\includegraphics[width=8cm]{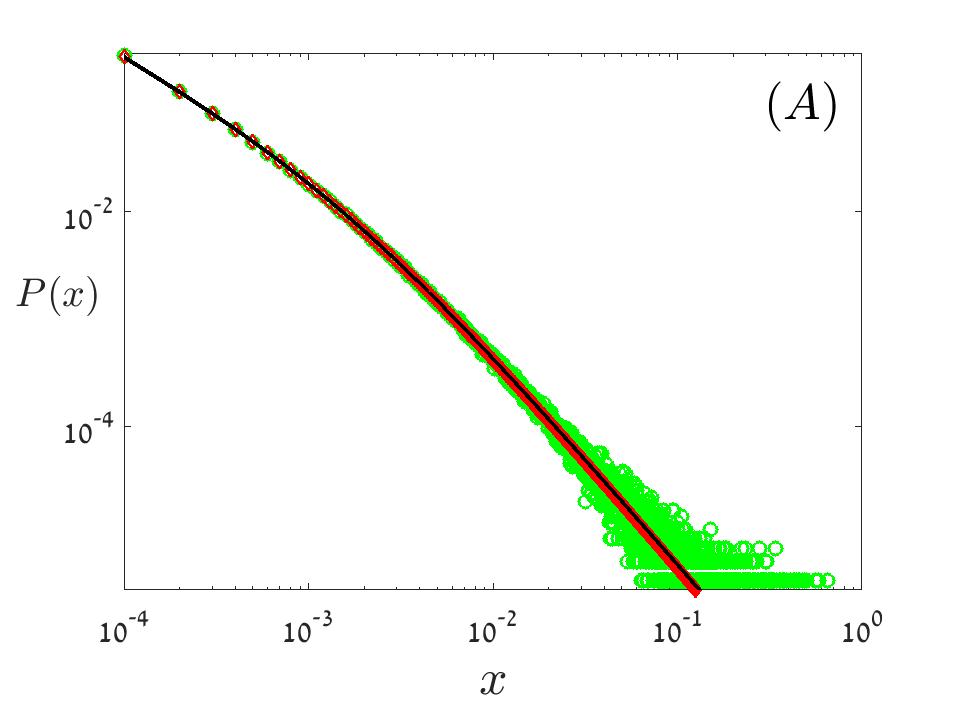}
\includegraphics[width=8cm]{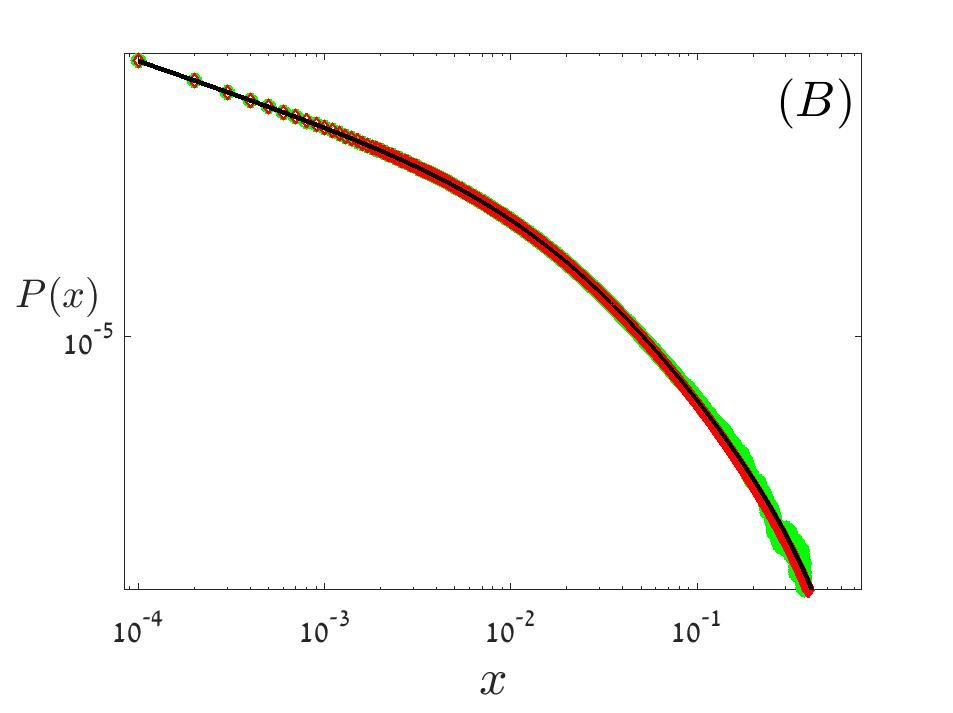}
\caption{The species abundance distribution, $P(x)$ as a function of $x$, for a time-averaged neutral model (model A) with environmental stochasticity and demographic noise. In both figures $N = 10^4$ and $\nu = 0.01$, and the results  are plotted using a double logarithmic scale. The outcomes of a Monte-Carlo simulation (green circles), numeric solution for the steady state of the master equations (see Appendix \ref{apc}) (red diamonds) and the analytic predictions of  Eq. (\ref{eq10}) (black line) are compared, and the fit is, again, quite good. In panel  (A) the results are depicted   for $\delta = 0.5$ and $\gamma = 0.5$, such that $\nu/g = 0.16$. In panel (B) $\delta = 0.25$ and $\gamma = 0.2$ so $\nu/g = 2$. In both panels the small $x$ behavior is $P \sim 1/x$, but in panel (A) this regime is very narrow since it requires $x \ll 1/G = 1.6 \cdot 10^{-3}$. The $Gx \gg 1$ behavior obeys a power law in panel (A), where the environmental stochasticity dominates ($\nu/g < 1/2$) and is dominated by an  exponential decay in panel (B), where the mutation losses are stronger. In these parameters, $N$ is not big enough to justify the use of the asymptotic value $\zeta = 1$. Instead, the value of $\overline{f}_+$ used in Eq. (\ref{eq10}) was obtained by measuring  the long-term average fraction of individuals in the plus state through the MC simulations.}\label{fig2}
\end{figure}

There is, again, a demographic regime: as long as $Gx \ll 1$ and $\theta x \ll 1$, $P(x) \sim 1/x$, as in Eq. (\ref{eq61}) above. When $Gx \gg 1$ one obtains,
\begin{equation} \label{eq16}
P(x) \sim  \frac{(1-x)^{\frac{\nu}{g} + \zeta -1}}{x^{\frac{\nu}{g}+\zeta + 1}} \exp \left(-\frac{\theta x/2}{1+Gx(1-x)} \right).
\end{equation}
This expression is very similar to (\ref{eq7}), and the only modification is the replacement of $\nu/g$ by $\nu/g+\zeta$ in the pre-exponential factor. This implies that the general analysis presented in the discussion of the two-species case still holds: for $\nu>2g$ the exponential truncation starts above  $x_c$ which is ${\cal O} (1/N)$ while for $\nu < 2g$, $x_c$ is  ${\cal O} (1)$.  The only qualitative difference between the multi-species and the two-species case appears in the $\nu<g/2$ regime, where the pre-exponential function grows above $x^* = 1/2 + \nu/(2g) + \zeta/2$. One may see a peak at finite value of $x$ only if $x^*<x_c$, a condition that translates to,
\begin{equation}
\nu < \frac{g(1-\zeta)}{2}.
\end{equation}
Therefore, when $N \to \infty$ and $\zeta \to 1$ there is no peak in the species abundance distribution (see Figure \ref{fig2}B, in comparison with Fig \ref{fig1}A). Since the decay is faster than $1/x^2$, the assumption $\zeta \to 1$ is self consistent.

In parallel with Figure \ref{fig1}, Figure \ref{fig2} demonstrates the ability of Eq. (\ref{eq10}) to fit both the numerical solution of the master equation and the outcomes of Monte-Carlo simulations. Note that, unlike the last section, here the agreement between the MC simulations and the numerics of the master equations is non trivial, since the master equations were built for a single species, assuming the ability to use an effective medium theory with one parameter, $f_+$.

\subsection{Model B}

 Now let us present the analysis of the multi-species version of model B presented above:  a community with time-averaged neutral dynamics, in which the competition is global and the dependence of the transition rates on the fitness is nonlinear.

 As before, we would like to reduce our analysis to a focal species and to encapsulate the effect of all other individuals by their average fitness $A$, defined as,
 \begin{equation}
 A = f_+ e^{2 \gamma} + (1-f_+)
 \end{equation}
 where, as before, $f_+$ measures the fraction (of all individuals that do not belong to the focal species) who are in the plus state. In parallel with our analysis of model A, we have multiplied the value of $\gamma$ by a factor of two, with respect to the two species game, in order to keep the overall strength of environmental stochasticity, $g$, to be $\gamma^2 \delta/2$.  as we shall see below,  here also the mean value of $f_+$ approaches $1/2 + \gamma \delta /4$.

 Naively, one would like to define  new transition probabilities for the focal species using $A$. Given the value of $\gamma$, these transition probabilities are,
 \begin{equation} \label{Bneutral1}
W_{n \to n+1} = (1-\nu)\frac{(1-x)x e^{2\gamma}}{(1-x)A+xe^{2\gamma}}, \qquad W_{n \to n-1} =  x \left(\nu + (1-\nu)\frac{(1-x)A}{(1-x)A+xe^{2\gamma}}\right).
\end{equation}
 From this point one may continue, as in model A, to derive the two coupled  master equations and the corresponding Fokker-Planck equations, from which an appropriate expression for $P(x)$ may be extracted.

When we did that, we discovered that the emerging formula for $P(x)$ does not fit the outcome of our Monte-Carlo simulations. It turned out that the origin of the problem is the $f_+$ fluctuations: since species flip continuously from the plus to the minus state and vice versa, the number of species in the plus state varies binomially. Accordingly, $f_+$, which is the number of species in the plus state times the average abundance of such a species - fluctuates such that $f_+ = \overline{f}_+ + \delta f_+$, where $\delta f_+$ is a random number taken, more or less, from a zero-mean gaussian distribution with width $\sigma \equiv \sqrt{Var(f_+})$. In general $\sigma \to 0$ as $N \to \infty$, but to fit the results of our simulations with finite $N$ we had to use this parameter. In model A this variance did not play any role, since the transition probabilities are linear in $f_+$ so the average $W_{n->n \pm 1}$ depends only of $ \overline{f}_+$. In contrast, here the nonlinearity of the $W$s compels one to take $f_+$ fluctuations into account.

Accordingly, we have implemented the procedure described above, replacing each of the $W$s of Equation (\ref{Bneutral1}) by $\tilde{W} = (1/2)W(\overline{f}_+ + \sigma)+(1/2)W(\overline{f}_+ - \sigma)$. It turns out that this is a decent approximation and there is no need to average the $W$s using the exact $f_+$ distribution. Doing that, we have expanded these average transition probabilities to a second order in $\gamma$, and used the approximated  $W$s to build the corresponding master and Fokker-Planck equations, in parallel with Eq. (\ref{eqc2}). After long and tedious calculations, the effective, one dimensional Fokker-Planck equation turns out to be of the same general form of (\ref{eq9}),
\begin{eqnarray} \label{eq9B}
&\ &\left[ x(1-x)\left(1 + G x(1-x)\right) P(x) \right]'' - \\ & \ & \left\lbrace \left[ Gx(1-x)(1-2 x) + x(1-x) N \left(  \gamma  (1-2 \overline{f}_+) + \gamma^2 \left[1-2x-4(1-x)\left(\overline{f}_+(1-\overline{f}_+)-\sigma^2\right)\right]\right) - \theta x \right] P \right \rbrace' = 0. \nonumber
\end{eqnarray}
Solving this equation for $P(x)$ one obtains,
\begin{equation} \label{eq10B}
P(x) = C \frac{(1-x)^{\theta}}{x(1-x) [1 + G x (1-x) ]^{\frac{\theta}{ 2}-\frac{1}{\delta}-\kappa}} \left[ \frac{1-(1-2x)\sqrt{\frac{G}{4+G}}}{1+(1-2x)\sqrt{\frac{G}{4+G}}} \right]^{\left(\frac{\theta}{2}-\zeta -\frac{1}{\delta}+\kappa \right)\sqrt{\frac{G}{4+G}}},
\end{equation}
where
\begin{equation}
\kappa = \frac{1}{\delta}\left(1-4\overline{f}_+(1-\overline{f}_+) + 4 \sigma^2\right).
 \end{equation}
Comparing (\ref{eq10B}) with Eq. (\ref{eq10}) one realizes that in the environmental-noise-controlled regime, $Gx(1-x) \gg 1$,
\begin{equation} \label{finalB}
P_{model \ B}(x) = (1-x)^{\frac{2}{\delta}} x^{2 \kappa} P_{model \ A}(x),
\end{equation}
since the value of $\kappa$ is typically small, while $1/\delta$ is a large factor in the interesting regime of strong stabilizing effect, the species richness of model B is typically larger than the species richness of model A for the same set of parameters.  When $\delta$ is large model A and model B have similar behavior and, if $G > \theta$ one expects that the SAD will be much wider than the Fisher log-series; this type of behavior was observed numerically in \cite{dean2017fluctuating}.

In the corresponding two-species model the stabilizing effect of the noise increases the chance of rare species to grow and of  common species  to shrink, thus stabilizing the $x=1/2$ state. Here we see that the same stabilizing mechanism causes an increase in the species richness, i.e., it decreases the mean abundance of a single species. Because $f_+ > 1/2$, in the multi-species model the mean fitness of a focal species is slightly smaller than the mean fitness of the community, and this effect almost cancels the noise-induced growth of rare species. Accordingly, the main impact of the stabilizing mechanism is to limit the growth of common species.

The expression (\ref{eq10B}) depends on the parameters $\overline{f}_+$  and $\sigma^2$.  When $N \to \infty$, these two parameters converge to $1/2 +\gamma \delta /2$ and zero, correspondingly.  For finite $N$ the situation is slightly more complicated. While in model A Eqs. (\ref{eq10}) for $P(x)$  and (\ref{eq13}) for $\zeta$ (that depends on $\overline{f}_+$) provide a closed form, here another equation has to be used in order to determine $\sigma^2$ in a self-consistent manner. 

To do that, we begin with the calculation of the species richness (SR).  As we show in Appendix \ref{apd}, the SR distribution is quite narrow with a peak at $1/\overline{x}$. Accordingly, we neglect temporal and system-to-system fluctuations in the species richness and approximate it by its peak value. Neglecting similar binomial fluctuations we assume that  $SR/2$ of the species are in the plus state and $SR/2$ in the minus state. The chance to find $f_+$ at a certain value, $\phi$, is thus the chance that the abundance of half of the species, which are in the plus state, sums up to $\phi$ while the sum of the abundances of the other half (who are in the minus state) is $1-\phi$. Now we make another approximation and assume that these two distributions, for the sum over the plus species and the minus species,  are identical, so $\sigma^2$ is not affected by the difference between $\overline{f}_+$ and $1/2$. Using the central limit theorem one has,
\begin{eqnarray} \label{extra}
&P&(f_+ = \phi) = C_1 \exp\left( -\frac{(\phi - \overline{x} \  {\rm SR} /2)^2}{ {\rm SR} \  {\rm Var}(x)} \right)\exp\left(- \frac{(1-\phi - \overline{x} \ {\rm SR} /2)^2}{ {\rm SR} \ {\rm Var}(x)} \right) \nonumber \\  &=& C_1 \exp\left(- \frac{(\phi - 1/2)^2}{{\rm Var}(x)/\overline{x}} \right)\exp\left( -\frac{(1-\phi - 1/2)^2}{ {\rm Var}(x)/\overline{x}} \right) =C_2 \exp\left( -\frac{2(\phi - 1/2)^2}{{\rm Var}(x)/\overline{x}} \right) ,
\end{eqnarray} 
where $C_1 $ and $C_2$  are normalization factors. As expected, Eq. (\ref{extra}) suggests a slightly wrong value, one half,  for $\overline{f}_+$. Nevertheless, it captured the leading contribution to $\sigma^2$, 
\begin{equation} \label{extra1}
\sigma^2 \equiv {\rm Var}(f_+) = \frac{{\rm Var}(x)}{4 \overline{x}}.
\end{equation}

\begin{figure}
\includegraphics[width=8cm]{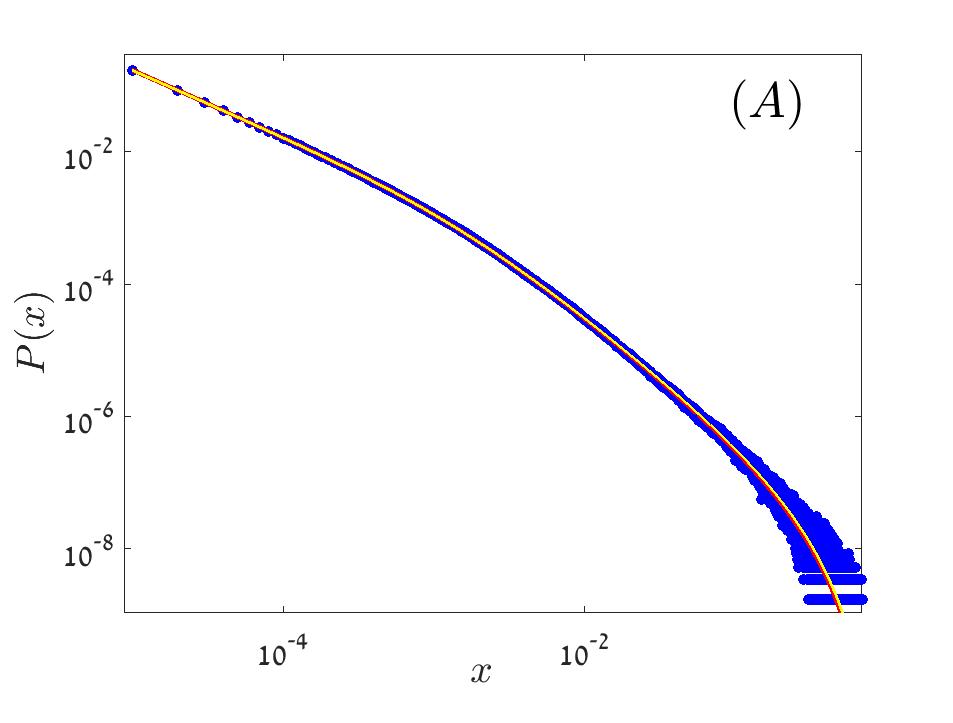}
\includegraphics[width=8cm]{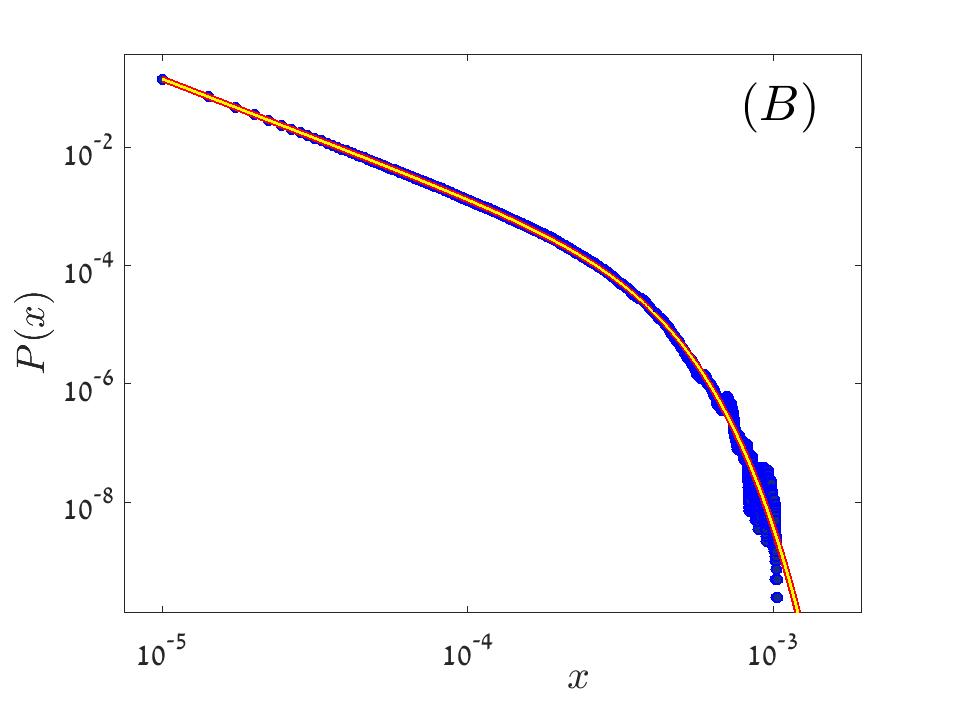}
\caption{The SAD, $P(x)$ vs. $x$, for model B. In both figures $N = 10^5$ and $\nu = 0.001$, and the results  are plotted using a double logarithmic scale. The outcomes of a Monte-Carlo simulation (blue circles), numeric solution for the steady state of the master equations (red line) and the analytic predictions of  Eq. (\ref{eq10B}) (yellow line) are compared. In panel  (A) the results are depicted   for $\delta = 0.5$ and $\gamma = \sqrt{0.025}$, such that $\nu/g = 0.16$. In panel (B) $\delta = 0.1$ and $\gamma = 0.05$ so $\nu/g = 8$. The values of $f_+$ and $\sigma$ were taken from the Monte-Carlo simulations. }\label{fig4}
\end{figure}

Eq. (\ref{extra1}) for $\sigma^2$, together with Eq. (\ref{eq10B}) for the distribution and Eq. (\ref{eq13}) for $\overline{f}_+$, provide a closed form from which $P(x)$ may be calculated iteratively by extracting $ \overline{x}$ and ${\rm Var}(x)$ from the distribution, plugging it in the expressions for $\overline{f}_+$ and $\sigma^2$ and iterating the process to convergence. This process allows us to fit the data in figure  \ref{fig4}.

These is, of course, a numerical alternative to this procedure: assuming a  distribution for $P_\pm(x)$ one may pick numbers from it until the sum reaches one, and calculate $f_+$. Iterating this for many times, a direct estimation of the mean and the variance of $f_+$ is obtained. We have verified our analytic approximation using this procedure, and the deviations (for the system parameters considered here) are smaller than $10 \%$.

\section{Discussion}

 The first neutral model, the neutral theory of molecular evolution, was suggested a few decades ago by Motoo Kimura \cite{kimura1985neutral}. By incorporating spatial effects (mainland-island dynamics), Stephen  Hubbell \cite{Hubbell2001unifiedNeutral} established the neutral model of biodiversity and biogeography. In both theories the diversity of a community reflects the balance between stochastic extinction and the emergence of new types via mutation, speciation, or  (on a local community in Hubbell's model) migration. The reproduction rate of all individuals is assumed to be equal at any time and the only driver of abundance fluctuations is demographic noise.

 In the immense body of literature published so far, neutral models are used in three distinct contexts: first, they serve as ultimate null models against which tests for selection or niche-based dynamics can be applied~\cite{nielsen2001statistical,leigh2007neutral}(though see \cite{gotelli2006null}). Second, these models describe the dynamics of all kinds of mutations and phenotypic variations that does not affect fitness (e.g., synonymous mutations). Third, even in systems like tropical trees or coral reefs one may argue that the very different species play, more or less, a neutral game, since the inferior species are already extinct, a mechanism known as emergent neutrality \cite{scheffer2006self,holt2006emergent,kessler2014neutral}.

 In this paper we have considered the simplest (and most important) neutral theory, the well-mixed model of Kimura which (without environmental noise) satisfies Ewens' sampling formula~\cite{ewens2012mathematical}. Under environmental variations that affect independently the relative fitness of species, such that all species still have the same time-average fitness, we provided here the average (over histories and states of the environment) SAD.

 Some will argue that our model does not deserve the title "neutral", since, for them, the concept of neutrality includes insensitivity of the system to the environmental conditions. However, it is clear that such an insensitivity is a matter of scale. Demographic and environmental stochasticity are the two extremes of the same phenomenon, namely, the stochastic effects of the environment on the fitness of a population: demographic noise
 is uncorrelated between different individuals, while the "environmental stochasticity" are those random variations that affect coherently an entire population. For us, neutrality means symmetry between (or equivalence of \cite{loreau2008species})  species, i.e., it corresponds to the assumption that the time-average fitness of all species is the same and that the dynamics is driven by (various kinds of) fluctuations.

 Previous works that dealt with this problems were focused on the dynamics of a single species with fluctuating growth rate, such that the time-averaged growth rate is $(-\nu)$ \cite{kessler2014neutral,fung2017species}. These works differ from the analysis presented here in two aspects: first, in our model the growth rate  (when a species is favored by the environment) decays with its fraction $x$. Second (and more important), by considering the  increase in the number of individuals in the plus state, which manifests itself in the value of $f_+ >1/2$. This second effect leads to an increased pressure on a focal species, hence the power-law decay [Eq. (\ref{eq16})] at large values of $x$ is characterized by an exponent which is larger than the exponent predicted for a two species game. In \cite{huerta2008population} the effect of the mutations on the growth rate of an existing species was neglected, and again the extra pressure due to $f_+ >1/2$ was not taken into account.

 In some circumstances, environmental stochasticity may act as a stabilizer of the community dynamics, increasing the chance of a new mutant to invade and decreasing chance of dominant species to grow.  This  phenomenon was pointed out by Chesson and coworkers \cite{chesson1981environmental,hatfield1997multispecies}, and our model B is similar to the lottery game considered by these authors, see~\cite{danino2016effect}. However, demographic noise and mutations were not taken into account, so their models did not allow for extinction and of course one cannot use them to study extinction-mutation equilibrium. Moreover, for neutral dynamics without demographic noise $\gamma$ cancels out from the steady state equations so the SAD depends only on $\delta$~\cite{hatfield1997multispecies,danino2016effect}: this happens because there is no other scale in the problem, but leads to the paradoxical result that the steady-state SAD is independent of the amplitude of environmental variations. When demographic stochasticity is taken into account, as we did here, the parameter $G = Ng$  sets the scale of environmental noise in terms of demographic stochasticity, and allows for a smooth transition between the purely demographic and the environmental models.

 In the original neutral model, with pure demographic noise and a Fisher log-series SAD, $P(x)$ decays like $1/x$ for $x \ll 1/(N \nu)$ and the decay is exponential above this point. In model A, the main effect of environmental stochasticity is to  allow for species with higher abundance; if environmental variations are  strong enough the exponential cutoff is replaced by a power-law decay as in Eq. (\ref{eq16}). This implies that in  such a system both the number and the abundance  of "hyperdominant" \cite{ter2013hyperdominance,slik2015estimate} species is larger, and the overall species richness is smaller, than in a system without environmental variations and the same speciation rate.  Recently, the heterogeneity of SADs obtained in the marine biosphere was shown to be greater than expected by a purely demographic neutral model \cite{connolly2014commonness} - this may be an indication for the effect of environmental variations. As species richness reflects a speciation-extinction balance, this observation is consistent with the results of previous works, where we showed that the time to absorption shrinks when environmental stochasticity  turned on and there is no mechanism that allows for noise induced stabilization~\cite{kessler2015neutral,danino2016stability,danino2017fixation}. 
The response of model B systems to environmental fluctuations is more intricate.  In a model without mutations and without demographic noise, the single species SAD peaks at $1/SR$~\cite{danino2016stability}, but this implies that such a system is  vulnerable to the invasion of a new species. The remnant of this behavior is the Beta-distribution-like function that multiply $P(x)$ of  model A to yield the SAD of model B in Eq. (\ref{finalB}).  When $\delta$ is large, model A and B behave similarly. However when $\delta$ is small and the stabilizing effect is strong, the SAD has a  strong cutoff at $x \sim \delta/2$ and the species richness increases substantially with respect to model A with the same parameters. Moreover, when $\delta < 2/\theta$ the species richness of model B will be even larger than the SR of the purely demographic neutral model that has a cutoff at $x=1/\theta$, as already demonstrated numerically in~\cite{danino2016stability} (Figure 5).

There are some limitations to our analysis: first, we assumed that  the size of the community $N$  is large, and that the number of species in the steady state is much larger than two (otherwise the mean-field approach failed, in particular the neglect  of the time-dependence of  $f_+$  becomes problematic).  Moreover, our approximations fail when $\delta$ becomes extremely large ($f_+ \to 1$), since in such a case the system reduces to a neutral model for all the plus state species, while the minus species simply go extinct. These limitations, of course, has nothing to do with the practical applications of the neutral model to empirical dynamics like those considered in \cite{kalyuzhny2015neutral,fung2016reproducing}. We believe that the theory presented here, when applied to  experiments and field data in population genetics and community ecology, may suggest many new insights into the processes that govern the composition of populations and communities. 

\section{\bf Acknowledgments} This research  was supported by the ISF-NRF Singapore joint research program (grant number 2669/17).
\bibliography{chesson_ref}

\appendix

\section{From master equation to Fokker-Planck equation: continuum approximation and boundary conditions} \label{apa}

The Fokker-Planck equations studied through this paper are obtained as a continuum approximation  of  an exact master equation. The justification for this procedure, and its limitations, were discussed in detain in \cite{kessler2007extinction}; in this appendix we provide a  few  comments that illustrate the  method used here,  with a particular emphasis on the boundary conditions. We stick to a simple system that allows us to demonstrate the problems and their solutions while keeping the algebra and calculus relatively  straightforward.

To begin, let us write down the master equation for a generic system with nearest neighbors transitions where the number of individuals is between one and $N$. If $W_{ n \pm 1 \to n}$ and $W_{n \to n}$ are the probabilities to jump into the state with $n$ individuals during one elementary step (after each elementary step, time is incremented by $1/N$), the master equation takes the form:
\begin{eqnarray} \label{eqa1}
P^{t+1/N}_1 &=& W_{1 \to 1} P^t_1 + W_{2 \to 1} P^t_2 \nonumber \\
P^{t+1/N}_n &=& W_{n \to n} P^t_n + W_{n+1 \to n} P^t_{n+1}+ W_{n-1 \to n} P^t_{n-1} \qquad 1< n < N\\
P^{t+1/N}_N &=& W_{N \to N} P^t_N + W_{N-1 \to N} P^t_{N-1}. \nonumber
\end{eqnarray}

In the steady state, $P^{t+1/N}_n =P^{t}_n$ for all $n$-s. In such a case the  set of equations (\ref{eqa1}) appears to provide $N$ equations for the $N$ unknown variables $P_n$. However, conservation of probability implies that the corresponding Markov matrix is singular, i.e., it admits a nontrivial eigenstate with zero eigenvalue. The missing constraint  is supplied by the normalization condition $\sum P_n =1$, and with this condition the solution is fully specified. This example may be generalized to include environmental noise, long-range hopping and so on.

Now let us discuss the transition to the continuum. The simplest way to make this approximation is to consider both $P$ and $W$ as functions of $x = n/N$, and to expand quantities like $P_{n+1} \to P(x+1/N)$ to second order in $1/N$. If it is possible to use this procedure for any value of $n$ (and this is not the case, see below) the equations for $P_1$ and $P_N$, which are not in the general form of all other equations, supply a  no-flux (Robin) boundary conditions at $x=0$ and $x=1$. As before, although one obtains a second order differential equation with two boundary conditions, the steady state is not specified completely  since the satisfaction of one boundary condition leads automatically to the satisfaction of the other one. The extra constraint is provided by normalization.

To examine the transition to continuum more closely, let us specify the transition probabilities.  As an example we take a two-species neutral model with pure demographic noise and  "reflecting" boundary conditions. At each step one individual is chosen at random to die and is replaced by an offspring of another, randomly chosen, individual. However, a singleton (the last individual that belongs to a certain species) cannot die. The corresponding transition probabilities are,

\begin{eqnarray} \label{eqa2}
W_{n-1 \to n} =  \frac{(n-1)(N-n+1)}{N(N-1)} &\qquad&   2 \le n \le N-1 \nonumber \\
W_{n+1 \to n} = \frac{(n+1)(N-n-1)}{N(N-1)}  &\qquad&   1 \le n \le  N-2 \nonumber \\
W_{n \to n} =  \left( 1- \frac{2n(N-n)}{N(N-1)} \right)  &\qquad&  2 \le n \le N-2  \\
W_{1 \to 1} =  ( 1- W_{1 \to 2}) &\qquad& W_{N-1 \to N-1} =  ( 1- W_{N-1 \to N-2}) \nonumber
\end{eqnarray}

Interestingly, for this model the steady state of the master equation (\ref{eqa1}) has a simple form,
\begin{equation} \label{eqa2A}
P_n = \frac{A}{n(N-n)}
\end{equation}
that satisfies both the master equation and the boundary condition. $A$ is determined by the normalization condition.

Plugging the transition probabilities in Eq. (\ref{eqa2}) into Eq. (\ref{eqa1}), the continuum equation is obtained by the set of replacements $n = xN$, $P_n \to P(x)$ and $P_{n \pm 1} \to  P(x) \pm P'(x)/N + P''(x)/2N^2$. The middle equation of  (\ref{eqa1}) is translated into,
\begin{equation} \label{eqa3}
\frac{dP(x,t)}{dt} = \frac{1}{N^2} \frac{\partial^2}{\partial x^2} \left(x(1-x)P(x)\right),
\end{equation}
and the steady state solution satisfies ${\dot P }  =0$, namely,
\begin{equation} \label{sta}
\frac{1}{N^2} \frac{\partial^2}{\partial x^2} \left(x(1-x)P(x)\right) = 0.
\end{equation}
The steady state solution of Eq. (\ref{eqa3}) has the general form,
\begin{equation} \label{sta1}
P(x) = \frac{A+Bx}{x(1-x)}.
\end{equation}
As explained above, one of the free constants $A$ and $B$ should be determined by (one of the) the boundary conditions, while the other allows for normalization. Comparing (\ref{sta1}) and (\ref{eqa2A}) one realizes that $B=0$ should be the correct answer, but the derivation of this result from the boundary conditions of the continuum differential equation is not trivial.

The problem (that has already been discussed in \cite{kessler2007extinction}) is that the continuum approximation itself may break close to $x=0$ and $x=1$. For example, in our case $P_1 \approx 2P_2$. Deriving the boundary condition from a continuum approximation, $P(2/N) = P(1/N) + P'(1/N)/N$, one finds $P'(1/N) = NP(1/N)/2$, but this in incompatible with $B=0$ in Eq. (\ref{sta1}) [$B=0$ implies $P'(1/N) = NP(1/N)$, without the factor 2]. This happens because the derivation of the boundary condition assumes that $P_n$ is smooth so the first derivative may be extracted from the difference between $P_1$ and $P_2$, but since the actual difference is a factor of $2$, the approximation fails and supplies the wrong boundary condition.

A way to solve this problem is to define another variable that will be smooth at the boundaries. For example,  the  quantity $Y = x(1-x) P$ undergoes a simple diffusion process so  Eq. (\ref{eqa3}) implies that at equilibrium $Y = A + Bx$, hence $Y'(x) = B$. The boundary condition  is translated to $Y(1/N) = Y(2/N)$, i.e., $Y'(1/N) =0$, and this implies $B=0$ as requested. However, we are not familiar with a  method that will allow us to produce a corresponding variable in more complicated scenarios.

The generic method, suggested in \cite{kessler2007extinction}, is to solve the difference (master) equation exactly at the vicinity of the boundary, and then to match this expression to  the solution of the differential (Fokker-Planck) equation in the bulk using the asymptotic matching technique. However, for the problems at hand this is a very complicated procedure and we have tried to avoid it.

Returning to the steady state equation (\ref{sta}), one notices that the constant $B$ is related to the first integration, i.e., $ \left(x(1-x)P(x)\right)'/N^2 = B$, so taking $B=0$ implies that after the first integration the remaining equation is still homogenous.  This is not an incident: it happens since the original problem satisfies  \emph{detailed balance}: $P_n W_{n \to n+1} = P_{n+1} W_{n+1 \to n}$, i.e.,  the probability flux between each pair of neighboring states is zero.

The detailed balance condition must hold in the steady state of Markov  chains, by induction from $P_1$. Accordingly, in any one-dimensional Fokker-Planck equation with the general form $[A(x)P(x)]'' + [B(x)P(x)]'=0$ and reflecting boundary conditions  one should omit the first integration constant. In the next appendices we consider systems that may, in principle, allow for loops, but we map them to a one-dimensional system, so as long as our approximation holds, the detailed balance condition must be satisfied. As $N $ increases this approximation becomes better and better, since the relative width of the boundary zone approaches zero.  Accordingly, through this paper we implement this detailed balance approximation (namely, we drop the first integration constant). The fits of our results to the numerical solutions of the master equations indicate that this is indeed a decent approximation.

\section{Fokker-Planck equation for the two-species model with one-way mutations} \label{apb}

In this appendix we derive the effective one-dimensional Fokker-Planck equation for a model with two species (types) $A$ and $B$, with both demographic and environmental stochasticity, and with one-sided mutations (an offspring of $A$ may mutate into  $B$, but an offspring of $B$ is always  a $B$), as  described in section \ref{sec2} of the main text.

To begin, let us introduce two quantities, $P^t_{n,+}$, the chance of finding the system with $n$ A-type individuals in the  $(+\gamma)$ state at time  $t$  and $P^t_{n,-}$, the chance of finding the system in the $(-\gamma)$ state with $n$ A-type individuals. The time evolution (time is incremented by $1/N$ after each elementary step) of $P_{n,\pm}$ is governed by the two coupled master equations:

\begin{eqnarray} \label{eqb1}
P^{t+1/N}_{n,+} &=&  P^{t}_{n+1,+} W^{++}_{n+1 \to n}  + P^{t}_{n-1,+} W^{++}_{n-1 \to n}  +P^{t}_{n,+}  W^{++}_{n \to n} \nonumber \\ &+& P^{t}_{n-1,-} W^{-+}_{n-1 \to n}  + P^{t}_{n+1,-} W^{-+}_{n+1 \to n}  + P^{t}_{n,-} W^{-+}_{n \to n}  \\
P^{t+1/N}_{n,-} &=&  P^{t}_{n+1,-} W^{--}_{n+1 \to n}  + P^{t}_{n-1,-} W^{--}_{n-1 \to n}  +P^{t}_{n,-}  W^{--}_{n \to n} \nonumber \\ &+& P^{t}_{n-1,+} W^{+-}_{n-1 \to n}  + P^{t}_{n+1,+} W^{+-}_{n+1 \to n}  + P^{t}_{n,+} W^{+-}_{n \to n}, \nonumber
\end{eqnarray}
where $W^{++}_{n-1 \to n}$, for example,  is the probability to increase the A-type population by one (from $n-1$ to $n$ individuals) while staying in the plus environment, and $W^{+-}_{n-1 \to n}$ is the chance that the environment switches from plus to minus and then the A-type population grows.

If the abundance of species $A$ is $n$, the chance of an interspecific duel for two, randomly picked individuals is $F_n = 2n(N-n)/N^2$ when $N \gg 1$. Using this notation we can write the transition probabilities as:

\begin{eqnarray} \label{eqb2}
W^{++}_{n+1 \to n} = \left(1-\frac{1}{\delta N}\right)  \left[ (1-\nu) F_{n + 1} \left(\frac{1}{2} - \frac{\gamma}{4}\right) + \nu \frac{n+1}{N} \right] &\qquad& W^{++}_{n-1 \to n} = \left(1-\frac{1}{\delta N}\right)  \left[ (1-\nu) F_{n -1} \left(\frac{1}{2} + \frac{\gamma}{4}\right) \right] \nonumber \\
W^{--}_{n + 1 \to n} = \left(1-\frac{1}{\delta N}\right) \left[(1-\nu) F_{n + 1} \left(\frac{1}{2} + \frac{\gamma}{4}\right) + \nu \frac{n+1}{N} \right] &\qquad& W^{--}_{n - 1 \to n} = \left(1-\frac{1}{\delta N}\right) \left[(1-\nu) F_{n - 1} \left(\frac{1}{2} - \frac{\gamma}{4}\right)  \right]  \nonumber \\
W^{-+}_{n+1 \to n} = \frac{1}{\delta N}  \left[ (1-\nu) F_{n + 1} \left(\frac{1}{2} - \frac{\gamma}{4}\right) + \nu \frac{n+1}{N} \right] &\qquad& W^{-+}_{n-1 \to n} = \frac{1}{\delta N}  \left[ (1-\nu) F_{n -1} \left(\frac{1}{2} + \frac{\gamma}{4}\right) \right]  \\
W^{+-}_{n + 1 \to n} = \frac{1}{\delta N} \left[(1-\nu) F_{n + 1} \left(\frac{1}{2} + \frac{\gamma}{4}\right) + \nu \frac{n+1}{N} \right] &\qquad& W^{+-}_{n - 1 \to n} = \frac{1}{\delta N}\left[(1-\nu) F_{n - 1} \left(\frac{1}{2} - \frac{\gamma}{4}\right)  \right] \nonumber \\
W^{++}_{n \to n} =  W^{--}_{n \to n}=   \left(1-\frac{1}{\delta N}\right) \left[ (1-\nu) (1-F_n) +\nu \left(1-\frac{n}{N}\right) \right] &\qquad&  W^{+-}_{n \to n} = W^{-+}_{n \to n}= \frac{1}{\delta N} \left[ (1-\nu) (1-F_n) +\nu \left(1-\frac{n}{N}\right) \right].  \nonumber
\end{eqnarray}

As explained, our system admits a single absorbing state at $n=0$ and the dynamics inevitably leads to the extinction of the A species, so we have to assume that, very rarely (on timescales that are much larger than the extinction time) a new A individual arrives and the game is played over and over again. If our interest is in the chance of A to have abundance $n$ conditioned on its existence in the system, we can merge together all the colonization-extinction periods. Colonizations are random in time, so the chance of a colonization during each state period is $1/2$.  This is equivalent to the use of the master equation (\ref{eqb1}) only for $n \ge 2$, while for $n=1$ the boundary equations are,
\begin{eqnarray} \label{eqb3}
P^{t+1/N}_{1,+} = P^t_{2,+} W^{++}_{2 \to 1} + P^t_{2,-} W^{-+}_{2 \to 1} + P^t_{1,+} W^{++}_{1 \to 1} + P^t_{1,-} W^{-+}_{1 \to 1} + \frac{1}{2} \left( [W^{++}_{1 \to 0} + W^{+-}_{1 \to 0}]P^{t}_{1,+} + [W^{--}_{1 \to 0} + W^{-+}_{1 \to 0}]P^{t}_{1,-} \right) \nonumber \\
P^{t+1/N}_{1,-} = P^t_{2,-} W^{--}_{2 \to 1} + P^t_{2,+} W^{+-}_{2 \to 1} + P^t_{1,-} W^{--}_{1 \to 1} + P^t_{1,+} W^{+-}_{1 \to 1} + \frac{1}{2} \left( [W^{++}_{1 \to 0} + W^{+-}_{1 \to 0}]P^{t}_{1,+} + [W^{--}_{1 \to 0} + W^{-+}_{1 \to 0}]P^{t}_{1,-} \right).
\end{eqnarray}

Eqs (\ref{eqb1}-\ref{eqb3}) define a linear equation
\begin{equation} \label{eqb4}
P^{t+1/N} = {\cal M} P^{t},
\end{equation}
where  $P^t$ is a $2N$ vector ($P_i = P_{n=i,+}$ for $i \le N$ and $P_{n=i-N,-}$ for $N < i \le 2N$)  and ${\cal M}$ is a $2N \times 2N$ Markov matrix. The steady state is the eigenvector of ${\cal M}$ with the (highest) eigenvalue $\lambda =  1$.  To obtain a solution for this steady state given a set of parameters that determine the elements of ${\cal M}$  we have solved numerically for this eigenvalue. As discussed in Appendix \ref{apa}, the overall scale of the steady state $P_n$-s is determined by the normalization condition.

Now we would like to develop a Fokker-Planck differential equation for this steady state distribution. Defining $P^+_{n}$  ($P^-_{n}$) as the chances to find the system with $n$ individuals in the plus (minus) state in a period between colonization and extinction, Eq. (\ref{eqb3}) takes the form,

\begin{eqnarray} \label{eqb5}
P^+_{n} &=&  P^+_{n+1} W^{++}_{n+1 \to n}  + P^+_{n-1} W^{++}_{n-1 \to n}  + P^+_{n}  W^{++}_{n \to n} \nonumber \\ &+& P^-_{n-1} W^{-+}_{n-1 \to n}  + P^-_{n+1} W^{-+}_{n+1 \to n}  + P^-_{n} W^{-+}_{n \to n}  \\  P^-_{n} &=&  P^-_{n+1} W^{--}_{n+1 \to n}  + P^-_{n-1} W^{--}_{n-1 \to n}  + P^-_{n}  W^{--}_{n \to n} \nonumber \\ &+& P^+_{n-1} W^{+-}_{n-1 \to n}  + P^+_{n+1} W^{+-}_{n+1 \to n}  + P^+_{n} W^{+-}_{n \to n}. \nonumber
\end{eqnarray}

Plugging (\ref{eqb2}) into (\ref{eqb5}) and using the definition $q \equiv 1/2 +\gamma/4$  (this is the parameter $q_A$, introduced in Section \ref{sec2},  in the plus state):
\begin{eqnarray} \label{eqb6}
P^+_n &=& \left( 1-\frac{1}{N \delta} \right) \left\lbrace (1-\nu)  \left( q F_{n-1}  P^+_{n-1}  + (1-q) F_{n+1}  P^+_{n+1}  + (1-F_n) P^+_n  \right) + \nu \left(\frac{n+1}{N} P^+_{n+1}+\frac{N-n}{N} P^+_n \right) \right\rbrace \nonumber \\ &+& \frac{ 1}{N \delta}\left\lbrace (1-\nu)  \left( (1-q) F_{n-1}  P^-_{n-1}  + q F_{n+1}  P^-_{n+1}  + (1-F_n) P^-_n  \right) + \nu \left(\frac{n+1}{N} P^-_{n+1}+\frac{N-n}{N} P^-_n \right) \right\rbrace  \\ P^-_n &=& \left( 1-\frac{1}{N \delta} \right) \left\lbrace (1-\nu)  \left( (1-q) F_{n-1}  P^-_{n-1}  + q F_{n+1}  P^-_{n+1}  + (1-F_n) P^-_n  \right) + \nu \left(\frac{n+1}{N} P^-_{n+1}+\frac{N-n}{N} P^-_n \right) \right\rbrace \nonumber \\ &+& \frac{ 1}{N \delta}\left\lbrace (1-\nu)  \left(  q F_{n-1}  P^+_{n-1}  + (1-q) F_{n+1}  P^+_{n+1}  + (1-F_n) P^+_n  \right) + \nu \left(\frac{n+1}{N} P^+_{n+1}+\frac{N-n}{N} P^+_n \right) \right\rbrace \nonumber
\end{eqnarray}

These two coupled  difference equations for $P^+$ and $P^-$ may be translated to another pair of coupled difference equations for their sum (which is the  chance to be at $n$, no matter what is the weather) and their difference,
\begin{equation} \label{eqb7}
P_n \equiv P^+_n + P^-_n  \qquad \Delta_n \equiv P^+_n - P^-_n.
\end{equation}

Defining $x \equiv n/N$ one may switch to the continuum limit, with $P_n \to P(x)$ and $P_{n \pm 1} \to P(x \pm 1/N)$. Expanding to second order in $1/N$, the emerging couple of steady-state differential  equations is,

\begin{eqnarray}  \label{eqb8}
(1-\nu) \left\lbrace \frac{1}{N} [x(1-x) \Delta]'' - \gamma [x(1-x) P]' \right\rbrace + \nu [ x \Delta]'  &=& \frac{2\Delta}{\delta \left( 1-\frac{2}{\delta N} \right) }   \\ \nonumber
(1-\nu) \left\lbrace \frac{1}{N}[x(1-x) P]''  - \gamma [x(1-x) \Delta] ' \right\rbrace  + \nu [ x P]'  &=&0
\end{eqnarray}

In what follows (and in the main text) we neglect the difference between $1-\nu$ and one, since in the relevant parameter regime $\nu$ is very small compared to one (otherwise one may replace, from now on, every $\nu$ by ${\tilde \nu} \equiv \nu/(1-\nu)$. In a very similar process where the rate of duels is 1 and the rate of mutations is $\nu$, this $(1-\nu)$ factor disappears).  Moreover, since we are interested in the large $N$, fixed $\delta$ limit,  $2/(\delta N) \ll 1$.

Dominant balance analysis (see discussion below) reveals that, for reasonably large $N$, the first and the third term in the upper equation of (\ref{eqb8}) are negligible. Accordingly,
\begin{equation} \label{eqb8a}
\Delta = \frac{\gamma \delta}{2} [x(1-x)P]'.
\end{equation}
When this expression is plugged  into the second equation one obtains an autonomous equation for $P$,
\begin{equation} \label{eqb9}
\left[ x(1-x) \left(\frac{1}{N} + gx(1-x) \right) P \right]'' - \left[ ( gx(1-x) (1-2x) - \nu x) P \right]' = 0,
\end{equation}
where $g \equiv \delta  \gamma^2/2$ is the strength of the environmental stochasticity. This equation and its solution for different parameter regimes are discussed in Section \ref{sec2} of the main text.

Our dominant balance analysis was based on numerical observations (solving numerically the Fokker-Planck equation and comparing the magnitude of different terms) but we can provide a few arguments for its  self consistency.

First, it is clear that in the demographic regime (i.e., $Gx \ll 1$) environmental fluctuations are negligible and the $\Delta$ terms are irrelevant, so the upper equation in (\ref{eqb8}) plays no role. By the same token, if the $P$ term in the upper equation is negligible in the large $N$ limit the solution is $\Delta=0$ and  the effect of environmental stochasticity disappears, so this term should be dominant when environmental variations are important.

 Let us define $Y(x) \equiv x(1-x)P$, so Eq. (\ref{eqb8a}) implies that $\Delta = (\gamma \delta/2) Y'$. Clearly, as long as (\ref{eqb9}) holds,
\begin{equation} \label{eqb10}
Y(x) = -\frac{1-x}{\nu} \left(\frac{1}{N} + gx(1-x)\right) Y'(x).
\end{equation} \label{eqb11}
and the dominant balance argument is consistent if, as $N \to \infty$, the two conditions,
\begin{equation}\label{eqb12}
\nu x \Delta  \ll \gamma Y \qquad  \frac{[x(1-x)\Delta]'}{N} \ll \gamma Y,
\end{equation}
or,
\begin{equation}\label{eqb13}
\frac{\delta \nu^2 x}{2} Y' \ll (1-x)\left(\frac{1}{N} + gx(1-x)\right) Y' \qquad   \frac{\delta \nu }{2N} [x(1-x) Y']' \ll (1-x)\left(\frac{1}{N} + gx(1-x)\right) Y',
\end{equation}
are satisfied.

When  $1/N \ll gx(1-x)$ the left condition is translated to $x \ll 1-\nu/\gamma$. On the other hand, if at large $N$ the third term balances the second, $\nu x \Delta  \sim \gamma Y$, one may plug it into the $\gamma [x(1-x)\Delta]'$ in the lower equation of (\ref{eqb8}) to find that this term is negligible with respect to the third one if $x > 1-\nu/\gamma$. Accordingly, in the regime where our dominant balance argument is wrong, environmental stochasticity is negligible.  Similarly, since the maximum value of $Y''/Y'$ is $\theta$,  the right condition in (\ref{eqb13})  holds when $x \ll 1-\nu^2/\gamma^2$, but if one assumes that the dominant balance is $\frac{[x(1-x)\Delta]'}{N} \sim \gamma Y$ and plug it into the lower equation of (\ref{eqb8}), the result is $Y \sim exp(-\theta x)$ and the effect of environmental noise vanishes for $\gamma^2 < \nu^2$, so we are back in the demographic regime.

\section{A Fokker-Planck equation for the multi-species model} \label{apc}

Unlike the two species game studied in Appendix \ref{apb}, here we consider the dynamics of a focal species in a multi-species environment. In a duel, an individual of the focal species may encounter an enemy with the same fitness (a neutral enemy), superior enemy (if the focal species is in the minus state) or inferior enemy (if it is in the plus state). As explained in the main text, we assume that the fraction of individuals in the plus state is fixed and equals to $f_+$. Accordingly, Eq. (\ref{eqb1}) still holds but the transition probabilities depend on the chance to find a neutral, superior or inferior enemy. If $\nu = 0$ these probabilities are,
\begin{eqnarray} \label{eqc1}
W^{++}_{n \pm 1 \to n} = \left(1-\frac{1}{\delta N}\right) F_{n \pm 1} \left[\frac{f_+}{2} +(1-f_+) \left(\frac{1}{2} \mp \frac{\gamma}{4}\right) \right] &\qquad& W^{--}_{n \pm 1 \to n} = \left(1-\frac{1}{\delta N}\right)  F_{n \pm 1} \left[\frac{1-f_+}{2}+f_+   \left(\frac{1}{2} \pm \frac{\gamma}{4}\right) \right]  \nonumber \\
W^{+-}_{n \pm 1 \to n} =  \frac{1}{\delta N} F_{n \pm 1} \left[\frac{1-f_+}{2}+f_+   \left(\frac{1}{2} \pm \frac{\gamma}{4}\right) \right]  &\qquad&  W^{-+}_{n \pm 1 \to n} =  \frac{1}{\delta N} F_{n \pm 1} \left[\frac{f_+}{2} +(1-f_+) \left(\frac{1}{2} \mp \frac{\gamma}{4}\right) \right]  \\
W^{++}_{n \to n} = W^{--}_{n \to n}= \left(1-\frac{1}{\delta N}\right) (1-F_n)  &\qquad&  W^{+-}_{n \to n} = W^{-+}_{n \to n}= \frac{1}{\delta N} (1-F_n).  \nonumber
\end{eqnarray}
As in Eqs. (\ref{eqb2}), when $\nu \neq 0$, each of these terms is multiplied by $(1-\nu)$, the quantity $\nu (n+1)/N$ is added to all the   $W_{n+1 \to n}$ terms and the quantity $\nu (1-n/N)$ is added to all the $W_{n \to n}$ terms.

Using the same boundary conditions (\ref{eqb3}), we can solve numerically for the steady state of the linear equation (\ref{eqb4}) using an iterative procedure: starting from  an initial value of $f_+$ we solved for the steady state, calculated (for this steady state) the new value of $f_+$ using the discrete version of Eq. (\ref{eq11}) and iterate this process until convergence.

Expanding equation (\ref{eqb5}), using the new $W$s, we obtained,
\begin{eqnarray}  \label{eqc2}
(1-\nu) \left\lbrace \frac{1}{N} [x(1-x) \Delta]'' - \gamma [x(1-x) (P+(1-2f_+)\Delta]' \right\rbrace + \nu [ x \Delta]'  &=& \frac{2\Delta}{\delta \left( 1-\frac{2}{\delta N} \right) }   \\ \nonumber
(1-\nu) \left\lbrace \frac{1}{N}[x(1-x) P]''  - \gamma [x(1-x) (\Delta + (1-2f_+)P] ' \right\rbrace  + \nu [ x P]'  &=&0
\end{eqnarray}
Using the  dominant balance argument and the approximations that we presented in the appendices above, the upper equation of (\ref{eqc2}) becomes,
\begin{equation} \label{eqc3}
\Delta = -\frac{\gamma \delta}{2} [x(1-x)P]'.
\end{equation}.
Plugging this expression for $\Delta$ into the lower equation one finds the effective Fokker-Planck equation for $P(x)$,
\begin{equation} \label{eqc4}
\left[ x(1-x) \left(\frac{1}{N} + gx(1-x) \right) P \right]'' - \left[ ( x(1-x) [g(1-2x)+ \gamma(1-2f_+)] - \nu x ) P \right]' = 0,
\end{equation}
which is Eq. (\ref{eq9}) of the main text.

\section{The species richness and its distribution} \label{apd}

This paper is focused on the species abundance distribution (SAD). In this appendix we would like to provide an expression for the overall species richness in the community given the SAD. To do that we implement standard tools which are relevant to any SAD, not only to those considered above.

We start from $P(x)$, the chance that a randomly chosen species has abundance $x$. Picking numbers at random from this distribution until their sum exceeds one, a possible instantaneous realization of the composition of the system is obtained. Defining the random variable
\begin{equation}
z_k = \sum_{j=1}^k x_j,
\end{equation}
one realizes that the cumulative distribution function (CDF) for the species richness is,
\begin{equation}
P({\rm SR} <k) = 1-P(z_k<1).
\end{equation}
The central limit  theorem suggests that $z_k$ is distributed like a  Gaussian random variable with mean $k \overline{x}$ and variance $k {\rm Var}(x)$. Accordingly, 
\begin{equation}
P({\rm SR} <k) = 1- \frac{1}{\sqrt{\pi}} \int_a^b \ dy \ e^{-y^2}.
\end{equation}
where $y \equiv (z_k-k\overline{x})/\sqrt{2k {\rm Var}(x)}$, $a =y(z_k =0)$ and $b =y(z_k =1)$. The distribution function for the species richness is the derivative of this CDF, and if $2 {\rm Var}(x) \ll \overline{x}$ (which is the common case), 
\begin{equation} \label{final}
P({\rm SR}=k)  = \frac{(\overline{x}  k+1) e^{-\frac{(\overline{x}  k-1)^2}{2 {\rm Var}(x) k}}}{2 \sqrt{2 \pi {\rm Var}(x) }  k^{3/2}}.
\end{equation}
Eq. (\ref{final}) is a slightly skewed Gaussian that peaks at $k = 1/\overline{x}$.

\end{document}